\documentstyle[12pt]{article}
 \makeatletter
\newcommand{\nn}{\nonumber}
\newcommand{\bd}{\begin{document}}
\newcommand{\ed}{\end{document}}
\newcommand{\bc}{\begin{center}}
\newcommand{\ec}{\end{center}}
\newcommand{\be}{\begin{eqnarray}}
\newcommand{\ee}{\end{eqnarray}}
\renewcommand{\thefootnote}{\alph{footnote}}
\newcommand{\se}{\section}
\newcommand{\sse}{\subsection}
\newcommand{\bi}{\bibitem}
\def\figcap{\section*{Figure Captions\markboth
     {FIGURECAPTIONS}{FIGURECAPTIONS}}\list
     {Figure \arabic{enumi}:\hfill}{\settowidth\labelwidth{Figure 999:}
     \leftmargin\labelwidth
     \advance\leftmargin\labelsep\usecounter{enumi}}}
\let\endfigcap\endlist \relax
\def\reflist{\section*{References\markboth
     {REFLIST}{REFLIST}}\list
     {[\arabic{enumi}]\hfill}{\settowidth\labelwidth{[999]}
     \leftmargin\labelwidth
     \advance\leftmargin\labelsep\usecounter{enumi}}}
\let\endreflist\endlist \relax
\begin{document}
\tolerance=10000 \baselineskip=7mm
\begin{titlepage}
 \vskip 0.5in
 \null
\begin{center}
 \vspace{.15in}
{\LARGE {\bf Study of $B_{s}\to (\eta,\eta',\phi)\ell\bar{\ell} $
Decays } }
\\ \vspace{1.0cm}
\par
 \vskip 2.1em
 {\large
  \begin{tabular}[t]{c}
{\bf C.Q. Geng and C.C. Liu}
\\
\\
{\sl Department of Physics, National Tsing Hua University}
\\  {\sl  $\ $ Hsinchu, Taiwan, Republic of China }
\\
\\
{\sl Theory Group, TRIUMF}
\\
{\sl  $\ $  4004 Wesbrook Mall, Vancouver, B.C. V6T 2A3, Canada}
\\
   \end{tabular}}
 \par \vskip 5.3em
\date{\today}
 {\Large\bf Abstract}
\end{center}
We study the rare decays of $B_{s} \to M_s\ell \bar{\ell}$
($M_s=\eta, \eta',\phi$ and $\ell=\nu,e,\mu,\tau$) in the standard
model. With the hadronic form factors evaluated in the light front
and constituent quark models, we estimate the decay branching
ratios, respectively. We also discuss the longitudinal lepton polarization 
and forward-backward asymmetries.

\end{titlepage}

\section{Introduction}
The inclusive flavor-changing neutral current (FCNC) process of
$B\to X_s \ell^+ \ell^-$ , which is suppressed and induced by
electroweak penguin and box diagrams in the standard model (SM),
has been recently observed by the Belle detector at the KEKB
$e^+e^-$ asymmetric-energy collider \cite{Belle1}
 with the branching ratio (BR) of $(6.1\pm1.4^{+1.3}_{-1.1})\times 10^{-6}$
 for di-lepton masses greater than $0.2\ GeV$,
 where $\ell$ is either an electron or a
muon and $X_s$ is a hadronic recoil system that contains a kaon.
The exclusive decay of $B\to K \ell^+ \ell^-$
has been  measured with the BR of
$(0.75^{+0.25}_{-0.21}\pm0.09)\times 10^{-6}$ \cite{Belle} and
$(0.78^{+0.24+0.11}_{-0.20-0.18})\times 10^{-6}$
\cite{BaBar,BaBar1} at the Belle
and Babar detectors
by using  $29.1\ fb^{-1}$
and $77.8\ fb^{-1}$ data samples,  which agree with the
theoretically estimated values
 \cite{Ali,CK,MNS,CG-PRD}, respectively.
 Experimental searches at the B-factories \cite{Belle,BaBar,BaBar1}
 for $B\to K^* \ell^+\ell^-$ are also close to the theoretical predicted ranges
 \cite{Ali,CK,MNS,CG-PRD}.
 At Babar, an excess of events over
background with estimated significance of $2.8\, \sigma$ has been
observed and $BR(B\to K^*\ell^+
\ell^-)=(1.68^{+0.68}_{-0.58}\pm0.28)\times 10^{-6}$ has been
obtained \cite{BaBar1}. It is clear that these FCNC rare decays
are important for not only testing the SM but probing new physics.

\begin{figure}[h]
\includegraphics{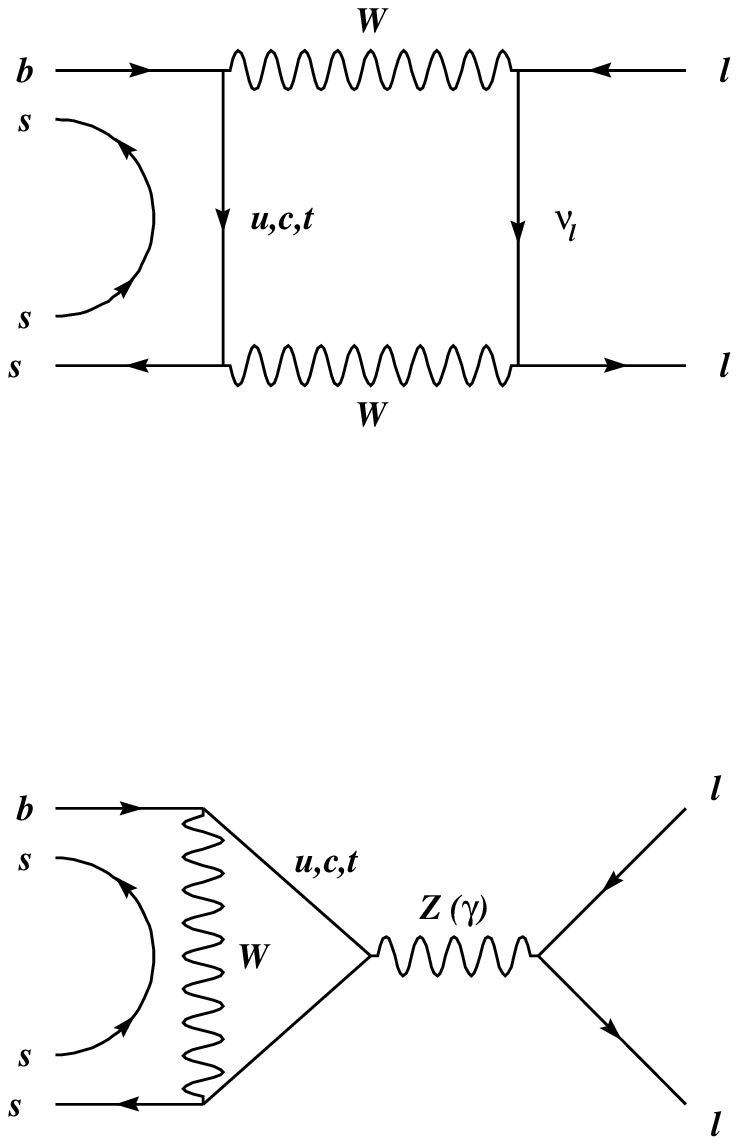} \vskip 13.cm\caption{One-loop diagrams for the
short-distance contribution in the standard model.}
\label{Feynman}
\end{figure}

In this paper, we study another type of the exclusive decays,
$B_s\to M_s\ell^+\ell^-$ ($M_s=\eta,\eta',\phi$), related to the
transition of $b\to s\ell^+\ell^-$ at the quark level as shown in
Figure \ref{Feynman}. The decays of $B_s\to\eta\ell^+\ell^-$ have
been estimated in Ref. \cite{eta1} and $B_s\to\phi\ell^+\ell^-$
have been explored in Refs. \cite{phi1,phi2}.
  Recently, the Collider
Detector at Fermilab (CDF) Collaboration has given an upper limit
of $BR(B_s \to \phi\mu^+ \mu^-)<6.7\times10^{-5}$ at a $95\%$
confidence level \cite{CDF:Bs}. In future, more data on the $B_s$
decays will be collected in various hadron colliders such as
CDF-II, HERA-B and ATLAS. To study the decay rates and branching
ratios, we need to calculate the transition form factors of the
vector, axial-vector and tensor currents.
There are many different candidates for this purpose, {\em e.g.}, lattice QCD
 \cite{9710057}, QCD sum rule \cite{sumrule}, and relativistic quark 
models. In this
work,  we use the framework of the light front quark model (LFQM) 
\cite{hwcw,tensor} to
evaluate  the form factors and compared with those in the constituent 
quark model (CQM)
\cite{DM:D62}.

This paper is organized as follows. In Sec. 2, we calculate the
form factors for $B_s \to \eta, \eta'$ and $\phi$ transitions in
the LFQM. We also present the results with those in the CQM. In
Sec. 3, we study the decay rates and asymmetries of $B_s \to P(V)
\ell^+\ell^-$  with $\ell=\nu,e,\mu,\tau$ and $P(V)=$ pseudoscalar
(vector) mesons. Our conclusions are given in Sec. 4.

\section{Form Factors}
\subsection{Effective Hamiltonians and Matrix Elements}

The contributions to the decays of $B_s\to M_s \ell\bar{\ell}~
(M_s=\eta,\eta',\phi)$ in the SM arise from the $W$-box and
$Z(\gamma)$- penguin diagrams as seen in Figure \ref{Feynman}. The
effective Hamiltonians of $b\to s\nu\bar{\nu}$ and $b\rightarrow s
\ell^+\ell^-$ are given by \cite{IL}
\be
    {\cal H}=\frac{G_{F}}{\sqrt{2}} \frac{\alpha }{2\pi \sin^{2}\theta _{W}}
     \lambda _{t} D\left( x_{t}\right) \bar{b}\gamma _{\mu }\left( 1-\gamma
      _{5}\right) s\bar{\nu }\gamma _{\mu}\left( 1-\gamma _{5}\right)
     \nu, \, \label{Hnunu}
\ee
and
\be
    {\cal H}=\frac{G_F\alpha \lambda _t}{\sqrt{2}\pi }\left[ C_8^{eff}
    \left( \mu \right) \bar{s}_L\gamma _{\mu }b_L\ \bar{\ell}
    \gamma ^{\mu }\ell +C_9\bar{s} _L\gamma _{\mu }b_L\
    \bar{\ell}\gamma ^{\mu }\gamma _5 \ell
    -\frac{ 2m_b C_7\left( \mu \right) }{q^2}\bar{s}_L i\sigma _{\mu \nu }
    q^{\nu }b_R\ \bar{\ell}\gamma ^{\mu }\ell\right]\,,
     \label{Hll}
\ee respectively,
 where $x_t \equiv m_t^2 /m_W^2$, $\lambda _{t}=V_{ts}^*V_{tb}$,
 $D(x_t)$ is the top-quark loop function \cite{geng1,Buras0}
  and $C_{8}(\mu )$, $C_{9}$ and $C_{7}(\mu )$ are Wilson coefficients 
(WCs) with their
explicit expressions given in Ref. \cite{Buras0} for the SM. We
note that $C_{9}$ is free of the $\mu $ scale. Besides the
short-distance (SD) contributions, the main effect on the decays
is from c\={c} resonant states such as $\Psi$ and $\Psi ^{\prime
}$, $i.e.$, the long-distance (LD) contributions. To including the
LD effect, in Eq. (\ref{Hll}) we
 replace $C_{8}(\mu)$ by $C_{8}^{eff}(\mu)$ \cite{CG-PRD,Buras0}, given by
\begin{equation}
 C_{8}^{eff}(\mu)=C_{8}( \mu) +\left( 3C_{1}\left( \mu \right) 
+C_{2}\left( \mu \right)
\right) \left( h\left( x,s\right)
+\frac{3}{\alpha }%
\sum_{j=\Psi ,\Psi ^{\prime }}k_{j}\frac{\pi \Gamma \left( j\rightarrow
 l^{+}l^{-}\right) M_{j}}{q^{2}-M_{j}^{2}+iM_{j}\Gamma _{j}}\right) \,, 
\label{effc8}
\end{equation}
 where we have neglected the small WCs, and $h(x,s)$ describes the 
one-loop matrix
elements of operators $O_{1}=\bar{s}_{\alpha }\gamma ^{\mu }P_{L}b_{\beta }\
\bar{c}_{\beta }\gamma _{\mu }P_{L}c_{\alpha }$
and $%
 O_{2}=\bar{s}\gamma ^{\mu }P_{L}b\ \bar{c}\gamma _{\mu }P_{L}c$ 
\cite{Buras0}, $M_{j}$
 ($\Gamma _{j}$) are the masses (widths) of intermediate states, and  
$k_{j}=-1/\left(
 3C_{1}\left( \mu \right) +C_{2}\left( \mu \right) \right) $ 
\cite{CG-PRD}. The WCs at
the scale of $\mu\sim m_b\sim4.8$ GeV are shown in Table \ref{wcs}.
\begin{table}[h]
   \caption{Wilson coefficients for $m_t=170$ GeV and $\mu=4.8$ GeV.}
   \label{wcs}
   \begin{center}
   \begin{tabular}{cccccc}
   \hline
  WC& $C_1$ & $C_2$ & $C_7$ & $C_8$ & $C_9$ \\ \hline
  &$-0.226$&$1.096$&$-0.305$&$4.186$&$-4.559$ \\ \hline
  \end{tabular}
   \end{center}
\end{table}

 To get the transition matrix elements of $B_s
\to M_s$ with various quark models, we parametrize them in terms
of the relevant form factors as follows:
\be
 \left\langle \eta_s(p_2)|\ V_\mu \ |B_s(p_1)\right\rangle
        &=& F_{+}(q^2) P_{\mu }+F_{-}(q^2) q_{\mu }\,,
         \nn \\
 \left\langle \eta_s(p_2) |\ T_{\mu\nu} q^\nu \ |B_s(p_1) \right\rangle
      &=& \frac{1}{m_{B_s}+m_{\eta(\eta')}}\left[ q^2P_{\mu }-\left( P\cdot
            q\right) q_{\mu }\right] F_T (q^2) \,, \nn \\
 \left\langle \phi(p_2,\epsilon ) |\ V_\mu \mp A_\mu \ |B_s(p_1)
         \right\rangle &=&\frac{1}{m_{B_s}+m_{\phi}}
         \left[-iV(q^2) \varepsilon _{\mu \nu \alpha \beta }
         \epsilon ^{*\nu }P^{\alpha }q^{\beta }
          \right.\nn \\ &&\pm A_0 (q^{2})\left(P\cdot q\right)
          \epsilon _{\mu }^{*} \pm A_+ (q^2) (\epsilon ^* \cdot p_1)P_{\mu }
           \nn \\
        && \left. \pm A_- (q^2) (\epsilon ^*\cdot p_1)q_{\mu }
        \right] \,, \nn \\
\left\langle  \phi(p_2,\epsilon ) |\ (T_{\mu\nu}
         \pm T^5_{\mu\nu})q^\nu \ |B_s(p_1) \right\rangle
         &=&-ig(q^2) \varepsilon _{\mu \nu \alpha \beta } \epsilon
         ^{*\nu }P^{\alpha }q^{\beta } \nonumber \\
      &&\pm a_0 (q^2) \left( P\cdot q\right) \left[ \epsilon _{\mu }^*
         -\frac{1}{q^2}\left( \epsilon^*\cdot q \right) q_{\mu }\right]
         \nn \\
      &&\pm a_{+}(q^2)(\epsilon ^*\cdot p_1)\left[ P_{\mu }
      -\frac{1}{q^2}\left( P\cdot p_1\right)q_{\mu }\right] \,,
\label{defo1} \ee
 where $m_M$ ($M=B_s$, $\eta$, $\eta'$ and
$\phi$) are the meson masses, $p_1 (p_2)$ is the momentum of the
initial (final) meson, $\epsilon$ is the polarization vector of
the vector meson $\phi$, $P=p_1+p_2$, $q=p_1-p_2$, $V_\mu=
\bar{q}_2\gamma _{\mu } q_1$, $A_\mu=\bar{q}_2\gamma _{\mu }
\gamma_5 q_1$, $T_{\mu\nu}=\bar{q}_2i\sigma _{\mu \nu } q_1$,
$T^5_{\mu\nu}=\bar{q}_2 i\sigma _{\mu \nu }\gamma_5 q_1$, and
$F_{\pm,T}$, $V$, $A_{0,\pm}$, $g$, and $a_{0,\pm}$ are the form
factors.

The daughter mesons in Eq. (\ref{defo1}) are in the bound state
$\bar{s}s$ with the physical masses, $m_{\eta}$, $m_{\eta'}$, and
$m_{\phi}$. In the transfer matrix elements of $B_s \to \eta_s$,
there are two sets of the form factors, $F_{\pm,T}$, corresponding
 $m_{\eta}$ and $m_{\eta'}$, respectively.

The pseudoscalar meson $\eta$ and $\eta'$ composed by the states
$u\bar{u}$, $d\bar{d}$ and $s\bar{s}$ can be expressed as the
mixtures of two orthogonal states $\eta_q$ and $\eta_s$,
given by $\eta_q=(u\bar{u}+d\bar{d})/\sqrt{2}$ and
$\eta_s=s\bar{s}$ \cite{feldmann}. Explicitly, one has that
\be
 \left (\matrix{\eta \cr \eta'}\right)\,=\,
             \left(\matrix{\cos{\theta} & -\sin{\theta} \cr
             \sin{\theta} & \phantom{-}\cos{\theta}} \right)
             \left (\matrix{\eta_q \cr \eta_s}\right) \,,
\ee with $\theta=(39.3^o\pm1.0^o)$ \cite{feldmann}.
 To study the decay rates of
$B_s \to (\eta,\eta')\ell\bar{\ell}$, we will first consider $B_s
\to \eta_s(m_{\eta})\ell\bar{\ell}$ and $B_s \to
\eta_s(m_{\eta'})\ell\bar{\ell}$ and then use
\be
   \Gamma(B_s\to\eta \ell\bar{\ell})=\sin ^2 \theta ~\Gamma
       [B_s \to \eta_s(m_{\eta}) \ell\bar{\ell}]
        \nn \\
   \Gamma(B_s\to\eta' \ell\bar{\ell})=\cos ^2 \theta ~\Gamma
       [B_s\to\eta_s(m_{\eta'}) \ell\bar{\ell}]\,.
\label{tran}
\ee

\subsection{Light Front Quark Model}
Since the calculations of the transition form factors in Eq.
(\ref{defo1}) belong to the nonperturbative regime, the
phenomenological quark models may be needed. One thing worthwhile
mentioning here is that all of form factors will be studied in the
time-like physical meson decay region of $0\leq q^2 \leq
(m_{B_s}-m_{\eta,\eta',\phi})^2$. As $q^2$ decreases,
corresponding to the increasing recoil momentum, we have to start
considering relativistic effects seriously. In particular, at the
maximum recoil point of $q^2=0$ where the final meson can be
highly relativistic, there is no reason to expect that the
non-relativistic quark model is still applicable. A consistent
treatment of the relativistic effects of the quark motion and spin
in a bound state is a main issue of the relativistic quark model.

The LFQM \cite{Ter,Chung} is the relativistic quark model in which
a consistent and fully relativistic treatment of quark spins and
the center-of-mass motion can be carried out \cite{Zhang}. The
LFQM has been applied to study the heavy-to-heavy and
heavy-to-light weak decay form factors in the timelike region
\cite{hwcw,time1}. These calculations are based on the observation
\cite{Dubin} that in the frame where the momentum transfer is
purely longitudinal, i.e., $q_\perp=0$, $q^2=q^+q^-$ covers the
entire range of momentum transfers. The price one has to pay is
that, besides the conventional valence-quark contribution, one
must also consider the non-valence configuration (or the so-called
$Z$ graph) arising from the quark-pair creation in the vacuum.
Unfortunately, a reliable way of estimating the $Z$ graph is still
lacking. However, the non-valence contribution vanishes if
$q^+=0$, and it is supposed to be unimportant for heavy-to-heavy
transitions \cite{hwcw}. In this paper, all of the values obtained
from the LFQM are based on the formulas in Refs.
\cite{hwcw,tensor}.

A meson bound state consisting of a heavy quark $q_1$ and an
antiquark $q_2$
 with total momentum $P$ and spin $S$ can be written as
\be
        |M(P, S, S_z)\rangle
              &=&\int  {dp_1^+d^2p_{1\bot}\over 2(2\pi)^3}
                       {dp_2^+d^2p_{2\bot}\over 2(2\pi)^3}
                ~2(2\pi)^3 \delta^3(\tilde
                P-\tilde p_1-\tilde p_2)~\nn\\
        &&\times \sum_{\lambda_1,\lambda_2}
                \Psi^{SS_z}(\tilde p_1,\tilde p_2,\lambda_1,\lambda_2)~
                |q_1(p_1,\lambda_1) \bar q_2(p_2,\lambda_2)\rangle,
\label{bound1} \ee where $p_1$ and $p_2$ are the on-mass-shell
light front momenta,
\be
        \tilde p=(p^+, p_\bot)~, \quad p_\bot = (p^1, p^2)~,
                \quad p^- = {m^2+p_\bot^2\over p^+},
\ee
with
\be
        && p^+_1=(1-x) P^+, \quad p^+_2=x P^+, \nn \\
        && p_{1\bot}=(1-x) P_\bot+k_\bot, \quad p_{2\bot}=x
        P_\bot-k_\bot\,.
\ee Here $(x,k_\perp)$ are the light-front relative momentum
variables, and $\vec{k}_\perp$ is the component of the internal
momentum $\vec{k}=(\vec{k}_\perp,k_z)$. The momentum-space
wave-function $\Psi^{SS_z}$ in Eq. (\ref{bound1}) can be expressed
as
\be
        \Psi^{SS_z}(\tilde p_1,\tilde p_2,\lambda_1,\lambda_2)
                = R^{SS_z}_{\lambda_1\lambda_2}(x,k_\bot)~ \phi(x, k_\bot),
 \label{momentumspace} \ee where $\phi(x,k_\bot)$ describes the momentum 
distribution of
the constituents in the bound state
\be
   \phi(x, k_\bot)={\cal N} \sqrt{dk_z \over dx} \exp\left( -
   \frac{\vec{k}^2}{2\omega ^2}\right) \label{wave}
\ee  with $\omega$ being the meson scale parameter and 
$R^{SS_z}_{\lambda_1\lambda_2}$
constructs a state of definite spin ($S,S_z$) out of light-front helicity
 ($\lambda_1,\lambda_2$) eigenstates. Explicitly, in practice it is more 
convenient to
use the covariant form for $R^{SS_z}_{\lambda_1\lambda_2}$ \cite{Jaus}:
\be
        R^{SS_z}_{\lambda_1\lambda_2}(x,k_\bot)
                ={\sqrt{p_1^+p_2^+}\over \sqrt{2} ~{\widetilde M_0}}
        ~\bar u(p_1,\lambda_1)\Gamma v(p_2,\lambda_2), \label{covariant}
\ee where
\be
\Gamma=\left\{
\begin{array}{ll} \gamma_5 & \mbox{(pseudoscalar
S=0)}\\ -\not{\!\hat{\varepsilon}}(S_z)+{\hat{\varepsilon}
        \cdot(p_1-p_2) \over M_0+m_1+m_2}
& \mbox{(vector S=1)} \end{array} \right. \ee and
\be
{\widetilde M_0} &\equiv &\sqrt{M_0^2-(m_1-m_2)^2} \label{M0} \ee
with
\be
        M_0^2={ m_1^2+k_\bot^2\over (1-x)}+{ m_2^2+k_\bot^2\over x}\,
\ee
 We normalize the meson state as
\be
        \langle M(P',S',S'_z)|M(P,S,S_z)\rangle = 2(2\pi)^3 P^+
        \delta^3(\tilde P'- \tilde P)\delta_{S'S}\delta_{S'_zS_z}~,
\label{waveno}
 \ee
so that the normalization condition of the momentum distribution
function can be obtained
\be
       \int {dx\,d^2k_\bot\over 2(2\pi)^3}~|\phi(x,k_\bot)|^2 = 1.
\label{momnor} \ee

We note that the form factors in Eq. (\ref{defo1}) depend on the
meson $(M=q_1\bar{q}_2$) wave functions $ \Psi_M(x, k_\perp)$. To
fix the parameters in the wave functions, one may use the meson
decay constants $f_M$, given by
\be
f_M &=& \sqrt{24} \int {dx\,d^2k_\perp\over 2(2\pi)^3}\,
        \Psi_M(x, k_\perp) \,{{\cal A}\over
        \sqrt{{\cal A}^2+k_\perp^2}}, \label{fp}
\ee where ${\cal A}=m_{q_1}x+m_{q_2}(1-x)$ with $m_{q_i}$ being
the composed quark masses.

In our numerical study, we use $f_{B_s}=200\ MeV$, and
$f_{\eta_s} = (1.34\pm 0.06)~f_{\pi}$ \cite{feldmann} which is the
decay constant of the constituent bound state $s\bar{s}$. By
putting these decay constants into Eq. (\ref{fp}), we get the
adapted $\omega$ value for the wave function in Eq. (\ref{wave}).
Explicitly, we find that $\omega _{B_s}=0.56$ and
$\omega_{\eta_s}=0.45$ GeV.

In Figure 2, we display the form factors as functions of
$s=q^2/M^2_{B_s}$ for the transitions of $B_s\to\eta_s$ (with
physical masses $m_{\eta, \eta'}$) and $B_s\to\phi$ with the
parameters in Table \ref{parameters}. In Table \ref{form1}, we
show the form factors of the pseudoscalar daughter meson with
$q^2=0$.
Since $F_+$ does not depend on the daughter meson mass, we find
that $F_+$ has the same value for both $m_{\eta}$ and $m_{\eta'}$,
whereas $F_-$ and $F_T$ are different.
  The form factors for $B_s \to \phi$
 including vector and tensor current matrix elements are given in Table 
\ref{form2}. We
 note that we have used the same scale parameter for the pseudoscalar and 
vetor mesons.
In Tables \ref{form1} and \ref{form2}, we have also listed
 the results in the CQM \cite{DM:D62}
 as comparisons.

We note that, as shown in Figure 2, there are deviations for the
form factors, in particular $F_T$, between the LFQM and CQM for
the region of large  $q^2$ values. As we know that in the LFQM the
form factors are calculated in all physical $q^2$ region, some
form factors, such as $F_T$, could increase as $q^2$ near the zero
recoil point since the matrix elements depend on the overlap
integral of the initial and final meson wave functions. While in
the CQM \cite{DM:D62}, all form factors have similar $q^2$
behaviors since they have the same double pole forms of

\be
{F_i(q^2)\over F_i(0)}&=& {1\over 1+\sigma_1s+\sigma_2s^2}
\label{Fit} \ee where $s=q^2/m_{B_s}^2$, $F_i(0)$ are the form
factors at $q^2=0$, and $\sigma _{1,2}$ are the fitted parameters.

\begin{table}[h]
    \caption{The parameters used in the calculation of the form factors 
(in GeV).}
   \label{parameters}
   \begin{center}
   \begin{tabular}{cccccccccc}
   \hline
   $m_{B_s}$&$m_{\eta}$&$m_{\eta'}$&$m_{\phi}$
&$m_b$&$m_s$&$f_{B_s}$&$f_{\eta_s}$&$w_{B_s}$&$w_{\eta_s}$
  \\ \hline
  $5.3696$& $0.5473$&$0.9578$&1.0194&$4.85$&$0.35$&$0.200$&
$0.183$&$0.56$&$0.45$ \\ \hline
  \end{tabular}
   \end{center}
\end{table}
\begin{table}[h]
   \caption{Form factors $F_{\pm,T}$ for $B_s \to \eta _s(m_{\eta},
   m_{\eta'})$ transitions at $q^{2}=0$ in the LFQM and CQM.}
   \label{form1}
   \begin{center}
   \begin{tabular}{cccccccc}
   \hline
   &\multicolumn{3}{c} {$B_s \to \eta_s(m_{\eta}$)} && \multicolumn{3}{c}
   {$B_s \to \eta_s(m_{\eta'})$}  \\   \hline
    &$F_+$&$F_-$&$F_T$&&$F_+$&$F_-$&$F_T$\\ \hline
   LFQM & $0.354$&$-0.360$&$-0.369$&&$0.354$&$-0.324$&$-0.404$ \\   \hline
   CQM & $0.357$&$-0.304$&$-0.365$&&$0.357$&$-0.304$&$-0.390$ \\ \hline
  \end{tabular}
   \end{center}
\end{table}

\begin{table}[h]
    \caption{Form factors $V,A_{0,\pm},g$ and $a_{0,+}$ for $B_s \to \phi$ 
for $B_s \to \phi$
    transitions at $q^{2}=0$ in the LFQM and CQM.}
   \label{form2}
   \begin{center}
  \begin{tabular}{cccccccc}
   \hline
    & \multicolumn{7}{c}{$B_s \to \phi$} \\ \hline
   &$V$&$A_0$&$A_+$&$A_-$&$g$&$a_0$&$a_+$ \\ \hline
   LFQM &$-0.440$&$-0.464$&$0.276$&$-0.295$&$0.377$&$0.374$&$-0.374$ \\
   \hline
   CQM &$-0.445$&$-0.506$&$0.310$&$-0.379$&$0.380$&$0.380$&$-0.380$ \\
   \hline
   \end{tabular}
   \end{center}
\end{table}

\section{Decay Rates and Asymmetries}

\subsection{Decay Rates}
Using Eqs. (\ref{Hnunu}), (\ref{Hll}), and (\ref{defo1}), we
obtain the decay amplitudes of $B_s \to (\eta_s,
\phi)\ell\bar{\ell}$ as follows:
\be
    {\cal M}(B_s\rightarrow \eta_s\nu \bar{\nu}) &=& \frac{G_{F}}{2\sqrt{2}}
       \frac{\alpha _{em}\lambda _{t}}{2\pi \sin ^{2}\theta
       _{W}}D(x_{t})\ F_{+}\ P_{\mu }\ \bar{\nu }\gamma ^{\mu }
       \left( 1-\gamma _{5}\right) \nu\,,
       ~~~~~~~~~~~~~~~~~~~~~~
       \label{Metanunu}\\
    {\cal M}(B_s\rightarrow \phi \nu \nu) &=&\frac{G_{F}}{2\sqrt{2}}
       \frac{\alpha _{em}\lambda _{t}}{2\pi \sin ^{2}\theta
       _{W}} D\left( x_{t}\right) \left\{ \frac{1}{m_{B_s}+m_{\phi}}\left[
       -iV\varepsilon _{\mu \nu \alpha \beta } \epsilon ^{*\nu }
       P^{\alpha }q^{\beta } \right.\right. \nn \\
       &&\left. \left. + A_{0}\left( P\cdot q \right)
       \epsilon _{\mu}^{*} + A_{+}\left( \epsilon \cdot P
       \right) P_{\mu}\right] \ \right\} \ \bar{\nu } \gamma ^{\mu }
       \left( 1-\gamma _{5}\right) \nu ~,
       \label{Mphinunu}\\
\nn\\
    {\cal M}(B_s\rightarrow \eta_s \ell^+\ell^-) &=& \frac{G_{F}\alpha_{em}
        \lambda _{t}}{2\sqrt{2} \pi }
        \left\{ \left[ \left( C_{8}^{eff}F_{+}
        -\frac{2m_{b}C_{7}F_{T}\left( q^{2}\right) }{m_{B_s}+m_{\eta(\eta')}}
        \right) P_{\mu }\right] \ \bar{\ell}\gamma ^{\mu }\ell \right.
        \nn \\
        &&\left. +\left[ C_{9}F_{+} P_{\mu }
        +C_{9}F_{-} q_{\mu }\right] \ \bar{\ell}
        \gamma ^{\mu }\gamma _{5}\ell\right\} \,,
        \label{Metall}
\\
    {\cal M}(B_s\rightarrow \phi \ell^+\ell^-)
           &=& \frac{G_{F}\alpha _{em}\lambda _{t}}
          {2\sqrt{2}\pi }\frac{2}{m_{B_s}} \nn \\
        &\times&\Bigg\{ ~\Bigg[ -i~G_V~
          \varepsilon _{\mu \nu \alpha \beta }~\epsilon ^{*\nu }P^{\alpha }
          q^{\beta } +G_A^0\left( P\cdot q\right) \epsilon _{\mu}^{*}
          +G_A^+\left( \epsilon ^{*}\cdot P\right) P_{\mu }\Bigg] \
          \bar{\ell}\gamma ^{\mu }\ell \nn \\
        &&+\Bigg[ -i~F_V~\varepsilon _
          {\mu \nu \alpha \beta }~\epsilon ^{*\nu }P^{\alpha }q^{\beta
          }+F_A^0\left( P\cdot q\right) \epsilon _{\mu
          }^{*}\nn \\
        &&~~+F_A^+\left( \epsilon ^{*}\cdot
          P\right) P_{\mu } +F_A^-\left( \epsilon ^{*}\cdot
          P\right) q_{\mu }\Bigg] \ \bar{\ell}\gamma ^{\mu }\gamma _{5}\ell
          ~\Bigg\} \,,\label{Mphill}
\ee where
the functions $G_V,~F_V,~G_A^0,~F_A^0,~G_A^+,~F_A^+$, and $F_A^-$
are defined by
\be
      G_V&=&\frac{C_8^{eff}~V}{2(1+\sqrt{r_{\phi}})}-\frac{C_7\hat{m}_bg}{s}\,,
      \quad F_V=\frac{C_9~V}{2(1+\sqrt{r_{\phi}})}\,,
       \nn \\
      G_A^0&=&\frac{C_8^{eff}~A_0}{2(1+\sqrt{r_{\phi}})}-\frac{C_7\hat{m}_ba_0}{s}\,,
      \quad  F_A^0=\frac{C_9~A_0}{2(1+\sqrt{r_{\phi}})} \,,
      \nn \\
      G_A^+&=&\frac{C_8^{eff}~A_+}{2(1+\sqrt{r_{\phi}})}-\frac{C_7\hat{m}_ba_+}{s}\,,
      \quad  F_A^+=\frac{C_9~A_+}{2(1+\sqrt{r_{\phi}})}\,,
       \nn \\
      F_A^-&=&\frac{C_9~A_-}{2(1+\sqrt{r_{\phi}})}\,,
\ee with $s=q^2/m_{B_s}^2$, $\hat{m}_b=m_b/m_{B_s}$ and
$r_{\phi}=m^2_{\phi}/m_{B_s}^2$.
From Eqs. (\ref{Hnunu}), (\ref{Hll}), (\ref{tran}) and
(\ref{Metanunu})-(\ref{Mphill}), the differential decay rates for
$B_s \to M_s \nu\bar{\nu}$ are found to be
\be
    \frac{d\Gamma \left( B_s\rightarrow M_s \nu \bar{\nu }\right) }{ds}
       &=&\frac{G_F^2|\lambda _t|^2
       \alpha _{em}^2|D\left( x_{t}\right) |^2m_{B_s}^5}{2^8\pi ^5
       \sin^4\theta_W}~\varphi_{M_s}^{1/2}~{\cal F}_{M_s} \,, \label{Rate1}
\ee where
\be
    {\cal F}_{\eta} &=& \sin^2\theta \times\varphi _{\eta}
        ~|F_+| ^2, \nn \\
    {\cal F}_{\eta'} &=& \cos^2\theta \times\varphi _{\eta'}
        ~|F_+| ^2,\nn \\
    {\cal F}_{\phi } &=& 3s\left[ (1-\sqrt{r}_{\phi })^2
|A_{0}|^2 +
           \frac{\varphi _{\phi }}{(1+\sqrt{r}_{\phi })^2}|V|
           ^2\right] \nn \\
        &+&\varphi _{\phi}\left[ \frac{(1-\sqrt{r}_{\phi })^2}{4r_{\phi }} |A_0|^2
           -\frac{s}{(1+\sqrt{r}_{\phi })^2}|V| ^2
           +\frac{\varphi _{\phi }}{4r_{\phi }(1+\sqrt{r}_{\phi })^2}|A_+|
           ^2 \right. \nn \\
        &+& \left.
           \frac{(1-r_{\phi}-s)(1-\sqrt{r}_{\phi})}{2r_{\phi}(1+\sqrt{r}_{\phi})}
           Re(A_0A_+^{*})\right].
           \ee
For $B_s \to M_s \ell^+\ell^-$ ($\ell=e, \mu, \tau)$, we get
\be
    \frac{d\Gamma \left( B_s\to M_s \ell^+\ell^-\right) }{ds}
        =\frac{G_F^2|\lambda _t|^2m_{B_s}^5\alpha _{em}^2}
        {3\cdot 2^9\pi ^5}~(1-\frac{4t}{s})^{\frac{1}{2}}~\varphi_{M_s} ^{1/2}
        \left[ \left(1+\frac{2t}
        {s}\right) \alpha_{M_s} +t~\delta_{M_s} \right] ~,
\label{Rate2} \ee where
\be
    &&t=m_{l}^{2}/m_{B_s}^{2}~, \quad
      r_{M_s}=m_{M_s}^{2}/m_{B_s}^{2}~, \quad
     \nonumber \\
    &&\varphi_{M_s}=\left( 1-r_{M_s}\right) ^{2}-2s\left( 1+r_{M_s}\right) +s^{2}~.
\ee
The formulas of $\alpha_{M_s}$ and $\delta_{M_s}$
are given by
\be
    \alpha_{\eta} &=& \sin^2\theta \times \varphi_{\eta} \left( |C_8^{eff}F_+
       -\frac{2C_7 F_T}{1+\sqrt{r_{\eta}}}|^2
       +|C_9F_+|^2\right) ~, \nonumber \\
    \delta_{\eta} &=& \sin^2\theta \times6|C_9|^2\{
       \left[2\left(1+r_{\eta}\right)-s\right] |F_+|^2+2\left( 1-r_{\eta}\right)
       Re(F_+F_-^*)+s|F_-|^2\}~,
       \\
       \nn\\
    \alpha_{\eta'} &=& \cos^2\theta \times\varphi_{\eta'} \left( |C_8^{eff}F_+
    -\frac{2C_7F_T}{1+\sqrt{r_{\eta'}}}|^2
       +|C_9F_+|^2\right) ~, \nonumber \\
    \delta_{\eta'} &=& \cos^2\theta \times6|C_9|^2\{
        \left[2\left(1+r_{\eta'}\right)-s\right] |F_+|^2+2\left( 1-r_{\eta'}\right)
       Re(F_{+}F_{-}^*)+s|F_{-}|^{2}\}~,
 \\
       \nn\\
    \alpha_{\phi} &=& 4s
    \left[ 3(1-r_{\phi})^2(|G_A^0|^2+|F_A^0|^2)+2\varphi_{\phi}(
    |G_V|^2+|F_V|^2)\right] \nn \\
    &&+\frac{\varphi_{\phi}}{r_{\phi}} \left[ (1-r_{\phi})^2
    (|G_A^0|^2+|F_A^0|^2)+\varphi_{\phi}(|G_A^+|^2+|F_A^+|^2)
    \right. \nn \\
    &&\left. +2(1-r_{\phi})(1-r_{\phi}-s)Re(G_A^0G_A^{+*}+F_A^0F_A^{+*}) \right] \nn \\
    \delta_{\phi} &=& -48 \varphi_{\phi}
    |F_V|^2-72(1-r_{\phi})^2|F_A^0|^2+\frac{6\left[2(1+r_{\phi})-s\right]}{r_{\phi}}\varphi_{\phi}
    |F_A^+|^2 \nn \\
    &&+\frac{6s}{r_{\phi}} \varphi_{\phi} |F_A^-|^2+\frac{12(1-r_{\phi})}{r_{\phi}}\varphi_{\phi}
    Re(F_A^0F_A^{+*}+F_A^0F_A^{-*}+F_A^+F_A^{-*})\,.
\ee

\begin{table}[h]
   \caption{Decay branching ratios of
$B_s \to (\eta, \eta', \phi) \nu\bar{\nu}$.}
   \label{br1}
   \begin{center}
   \begin{tabular}{lrr}
   \hline
   ~ & LFQM & CQM \\ \hline
   $10^{6}$Br$( B_s \to \eta \nu \bar{\nu} ) $&$2.34$&$2.17$ \\
   $10^{6}$Br$( B_s \to \eta' \nu\bar{\nu}) $&$2.52$&$2.38$ \\
   $10^{6}$Br$( B_s \to \phi \nu\bar{\nu}) $&$12.02$&$11.65$ \\
   \hline
   \end{tabular}
   \end{center}
\label{tablenu}
\end{table}

\begin{figure}[h]
\includegraphics{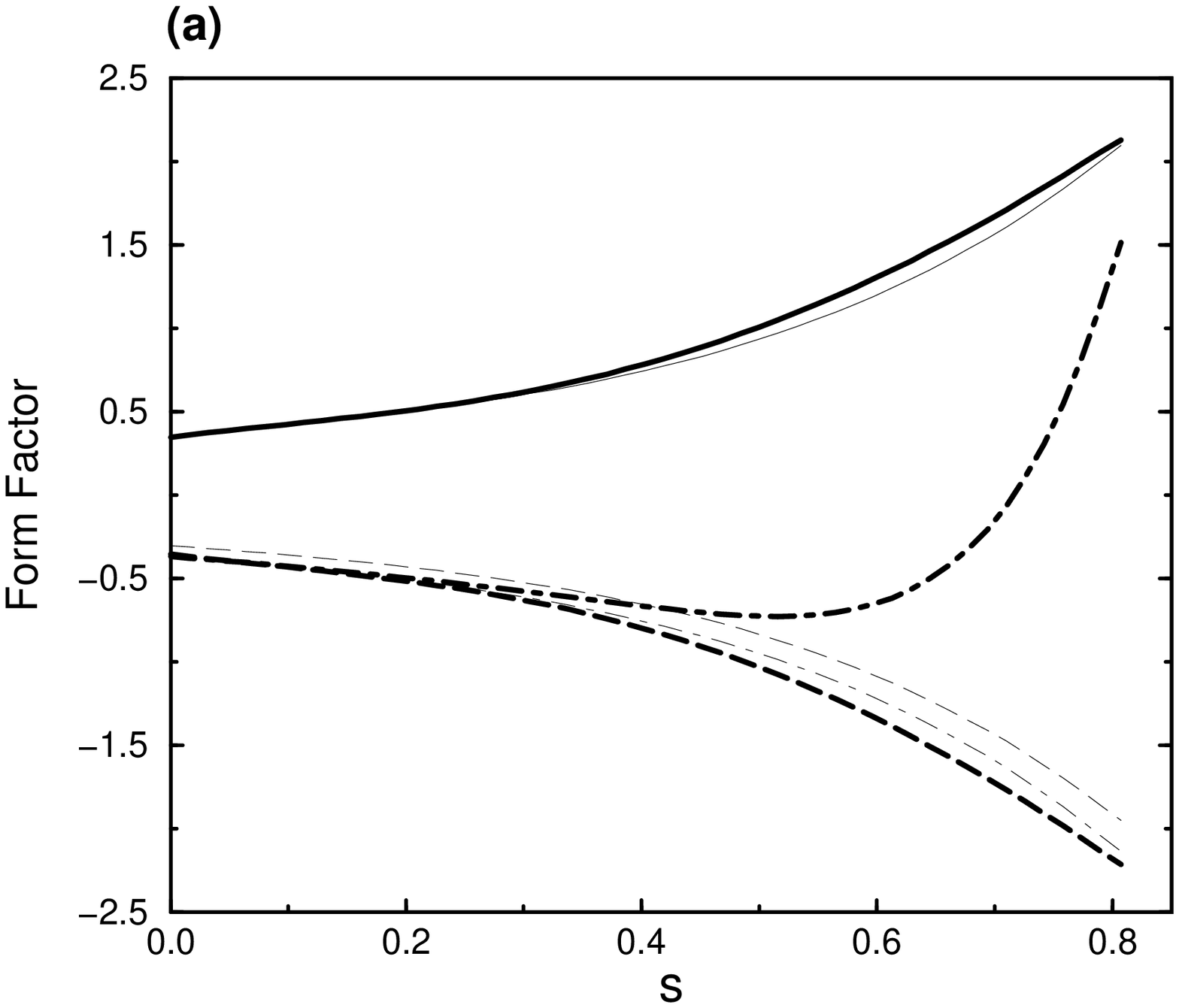}
\includegraphics{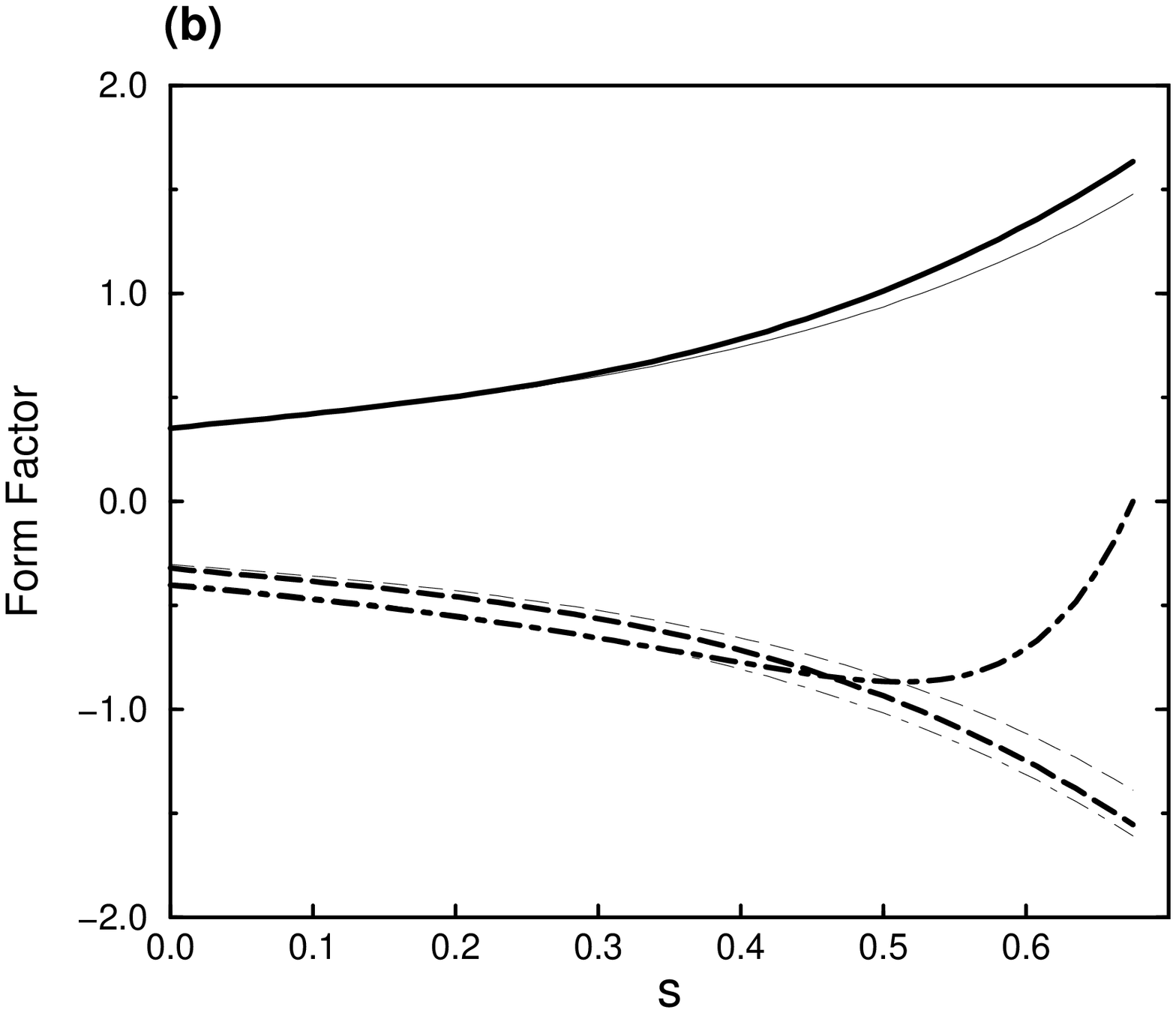}
\includegraphics{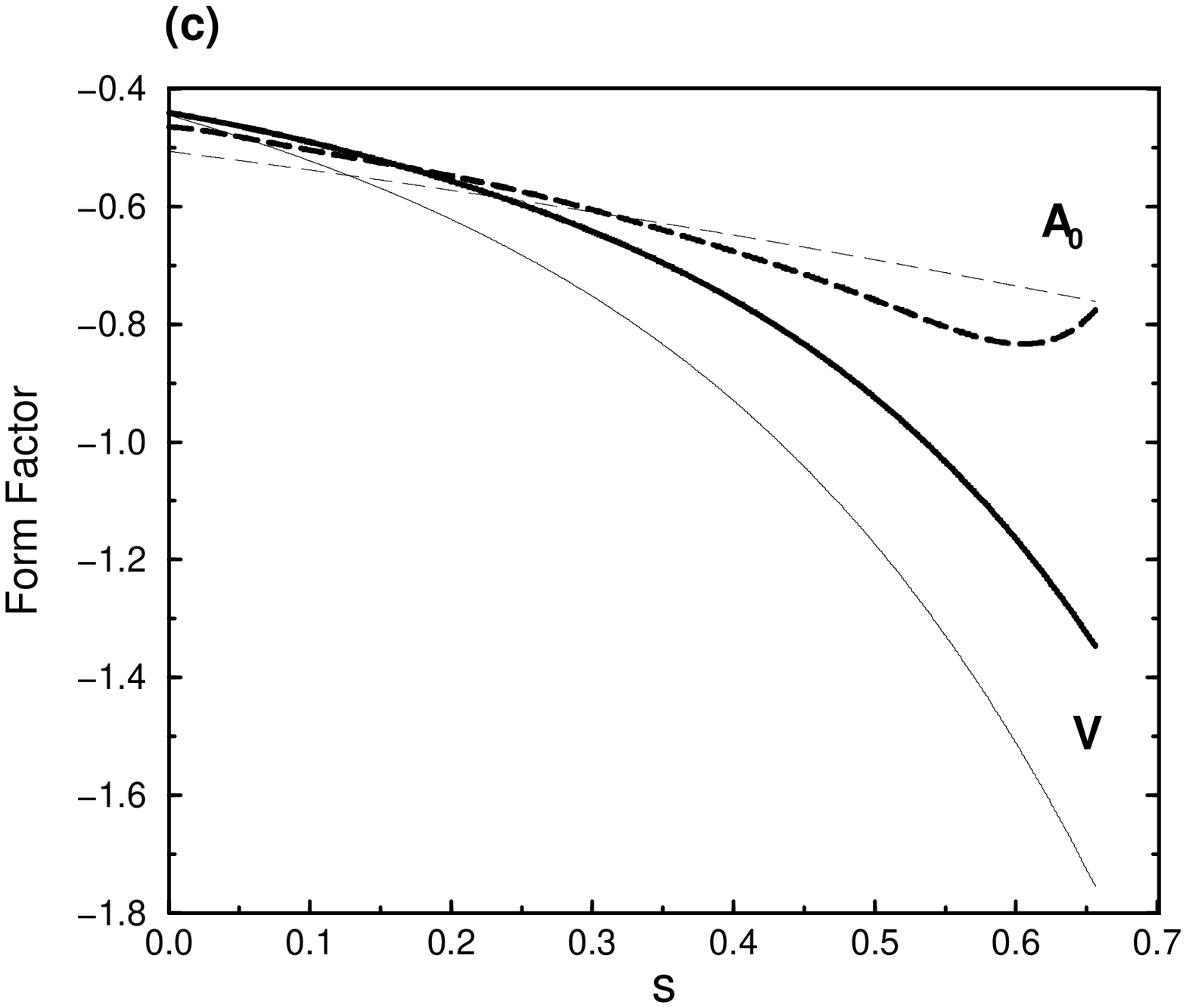}
\includegraphics{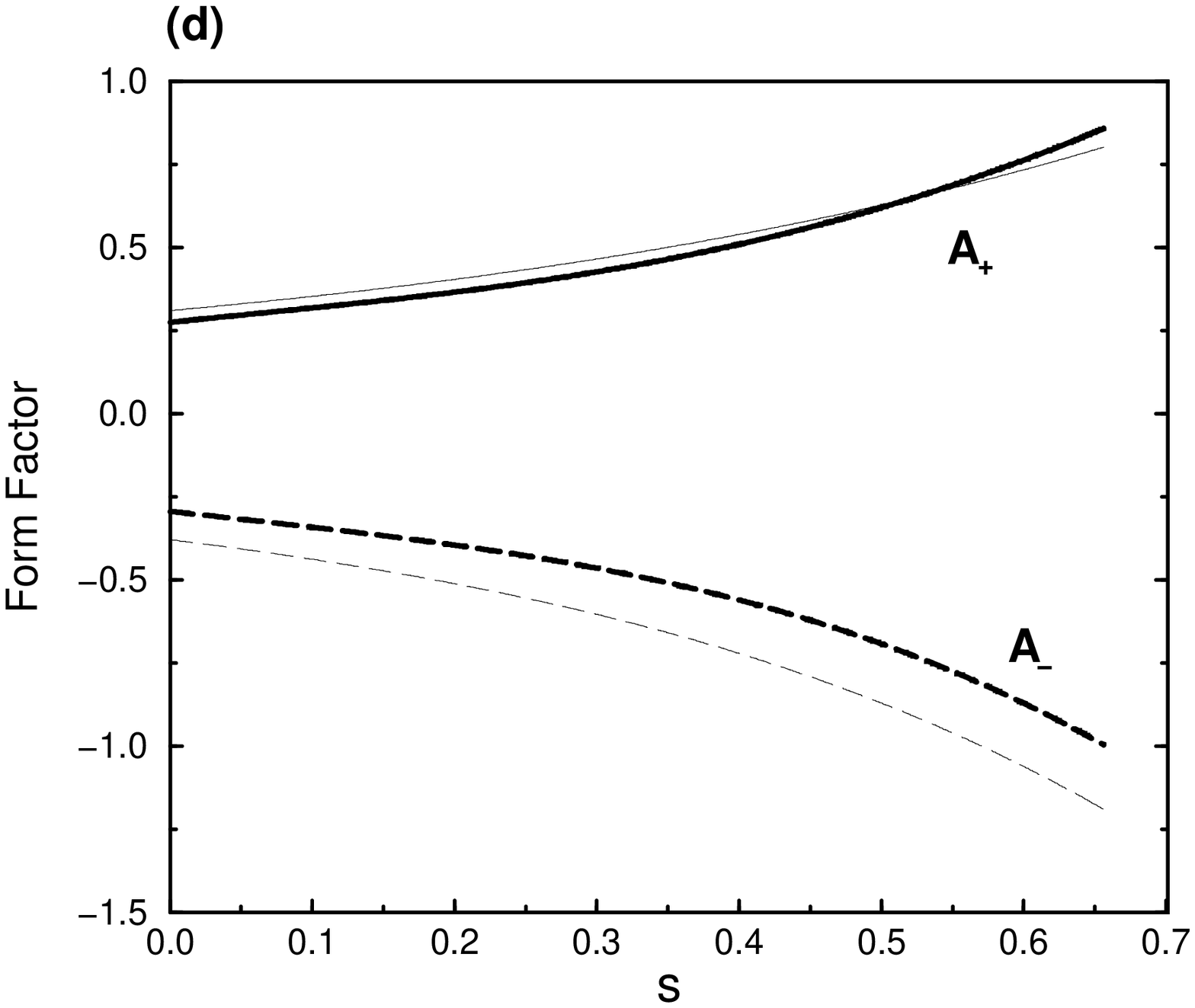}
\includegraphics{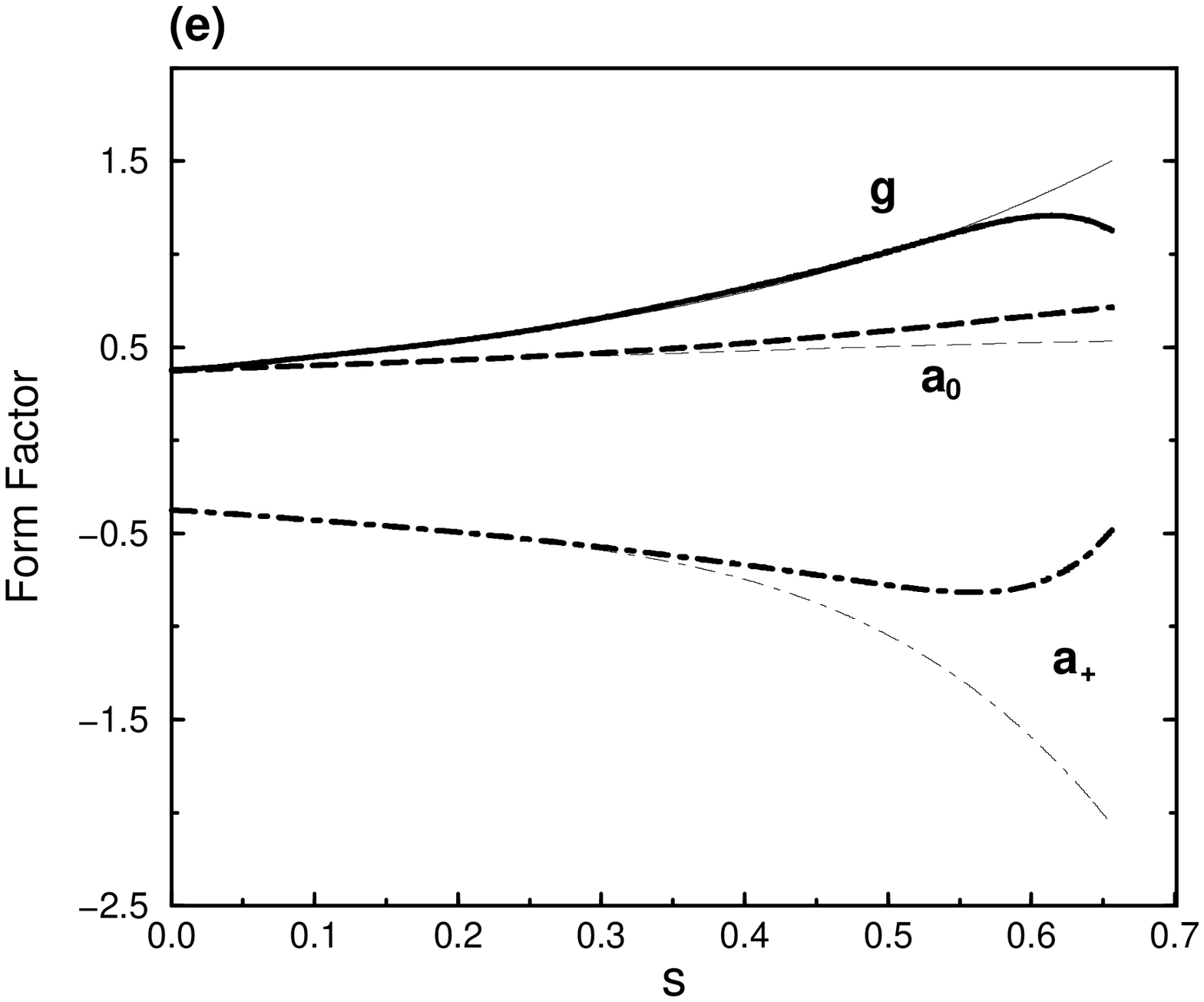}
\vskip 17.0cm \caption{ Form factors as functions of
$s=q^2/M^2_{B_s}$ for (a) $B_s\to \eta_s(m_{\eta})$  and (b)
$B_s\to\eta_s(m_\eta')$  with solid, dash, and dash-dot curves
representing $F_+$, $F_-$, and $F_T$, and (c,d,e) $B_s\to\phi$,
respectively.
The thick and thin lines stand for the results from the LFQM and
CQM, respectively.} \label{forms}
\end{figure}

\begin{figure}[h]
\includegraphics{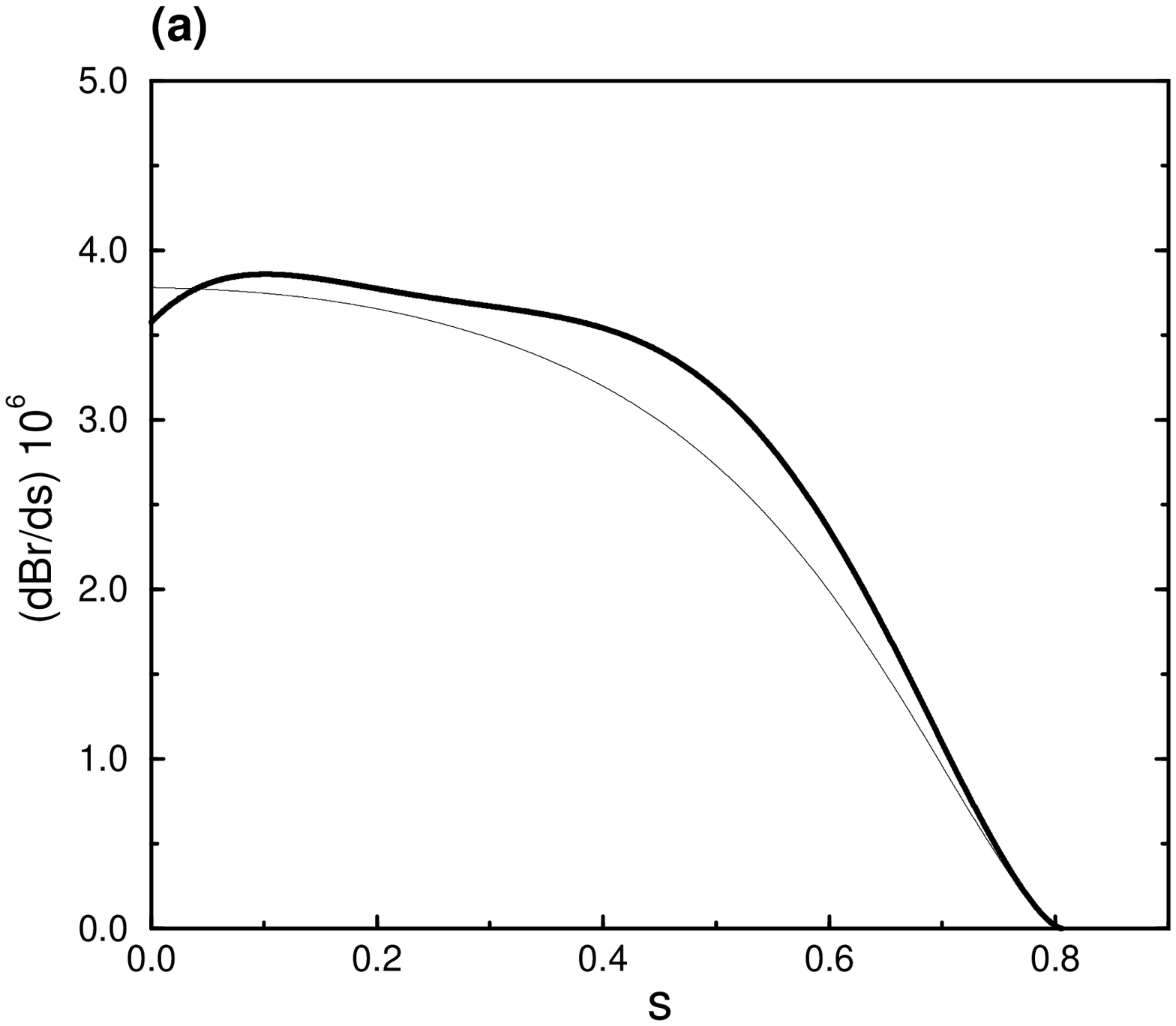}
\includegraphics{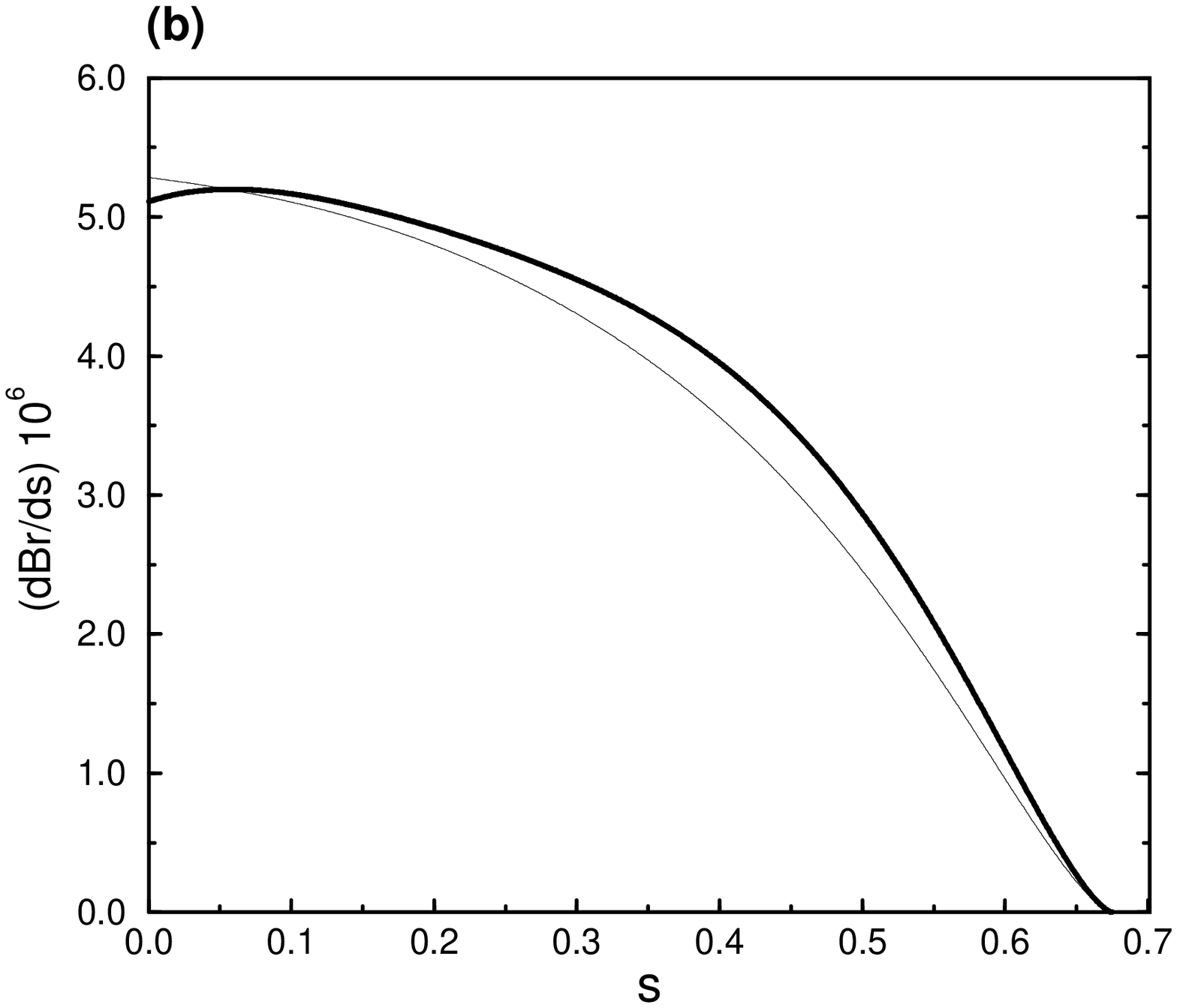}
\includegraphics{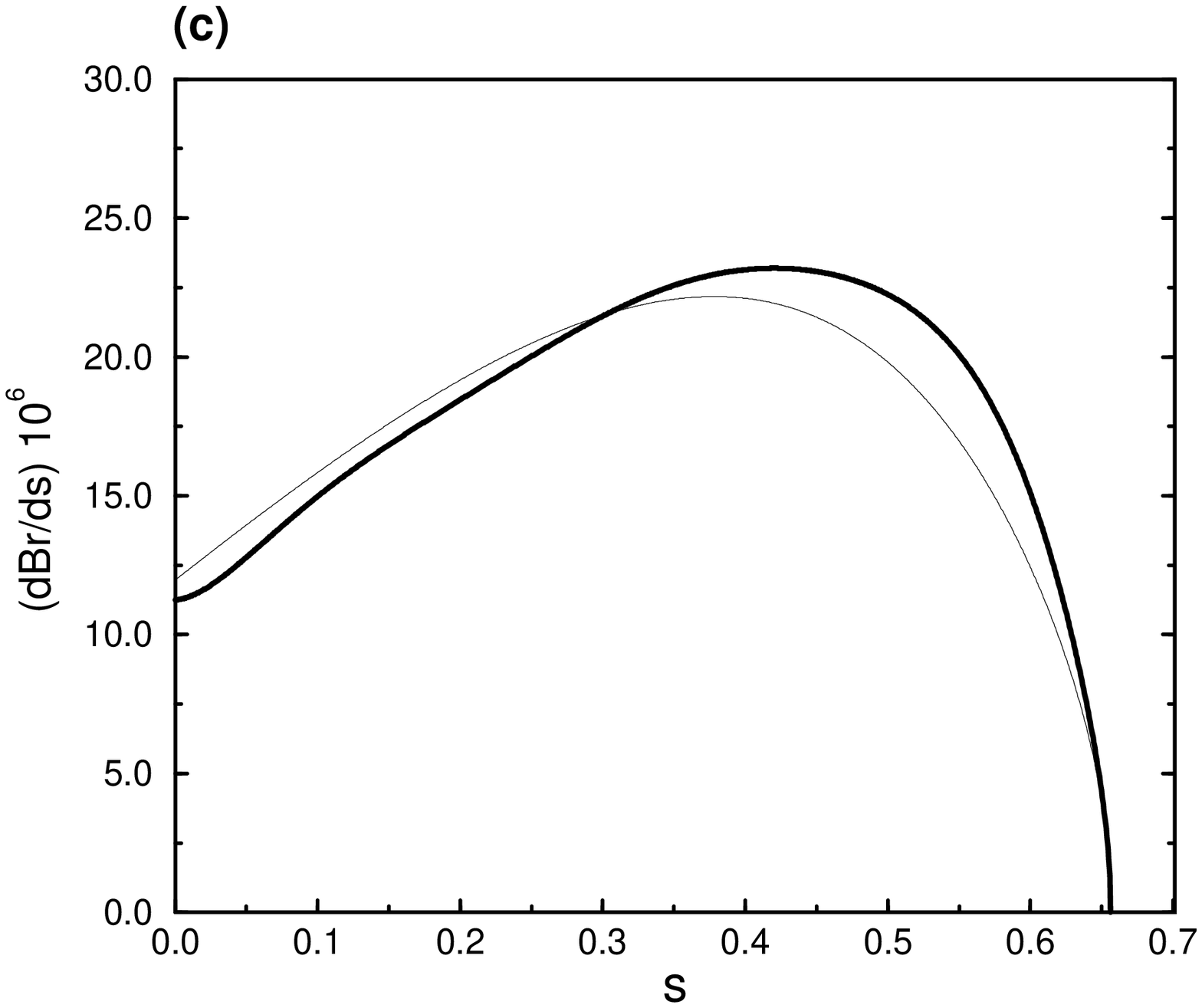}
\vskip 17.0cm \caption{ Differential decay branching ratios as a
function of $s=q^{2}/m_{B_s}^{2}$ for (a) $B_s\rightarrow \eta~
\nu\bar{\nu} $, (b) $B_s\rightarrow \eta' \nu\bar{\nu} $ and (c)
$B_s \to \phi ~\nu\bar{\nu}$. Legend is the same as Figure
\ref{forms}.
} \label{fig3}
\end{figure}

\begin{figure}[h]
\includegraphics{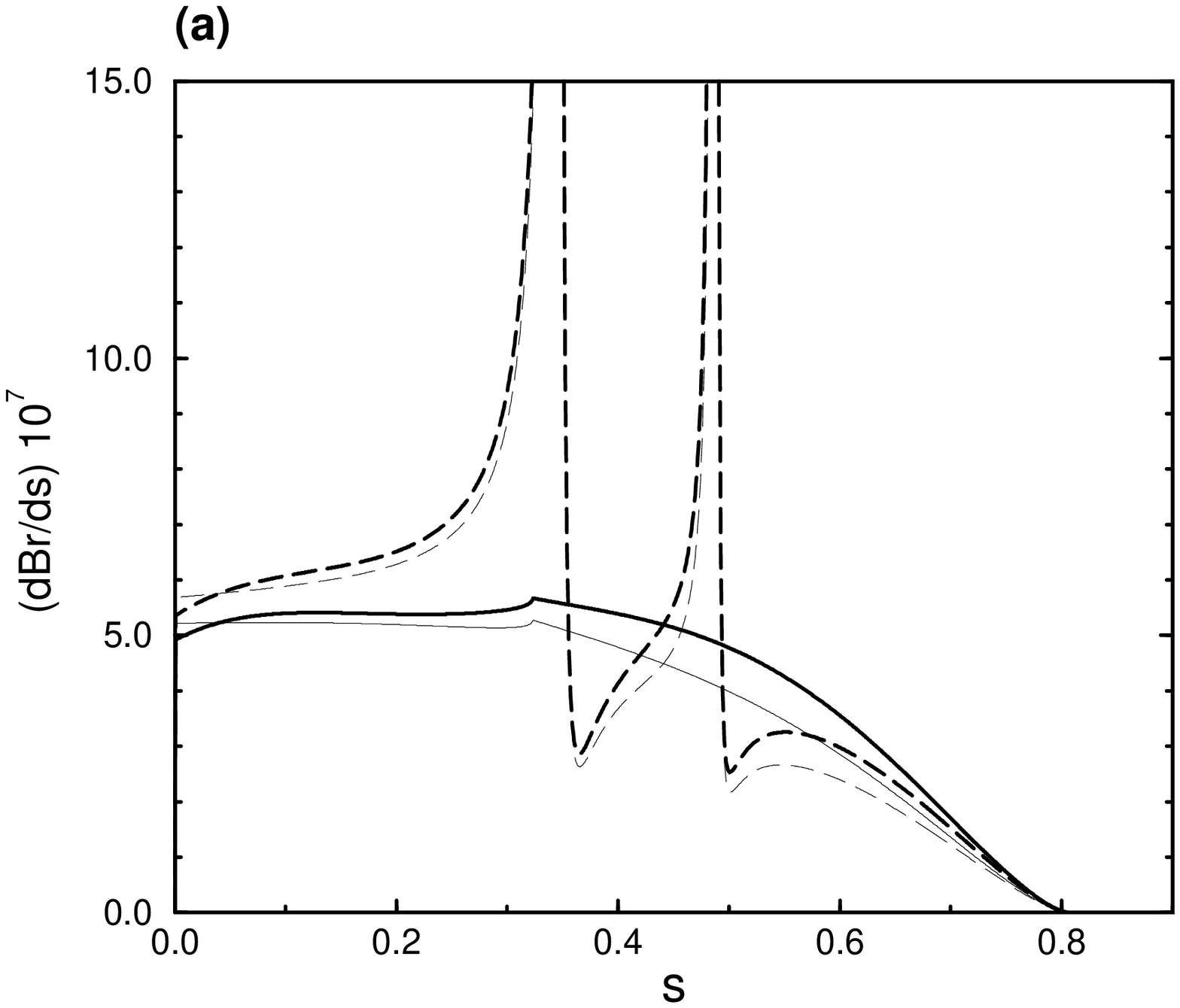}
\includegraphics{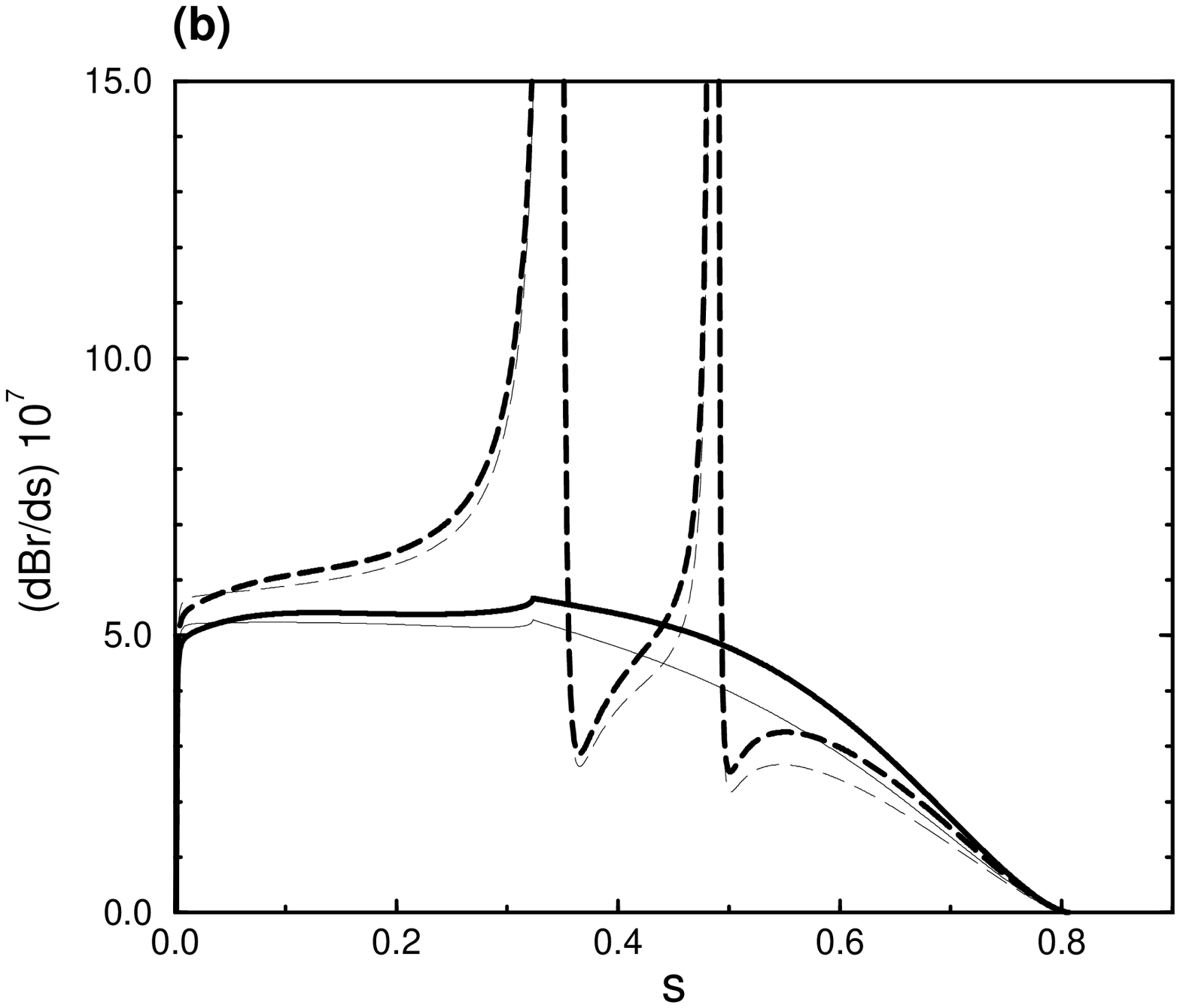}
\includegraphics{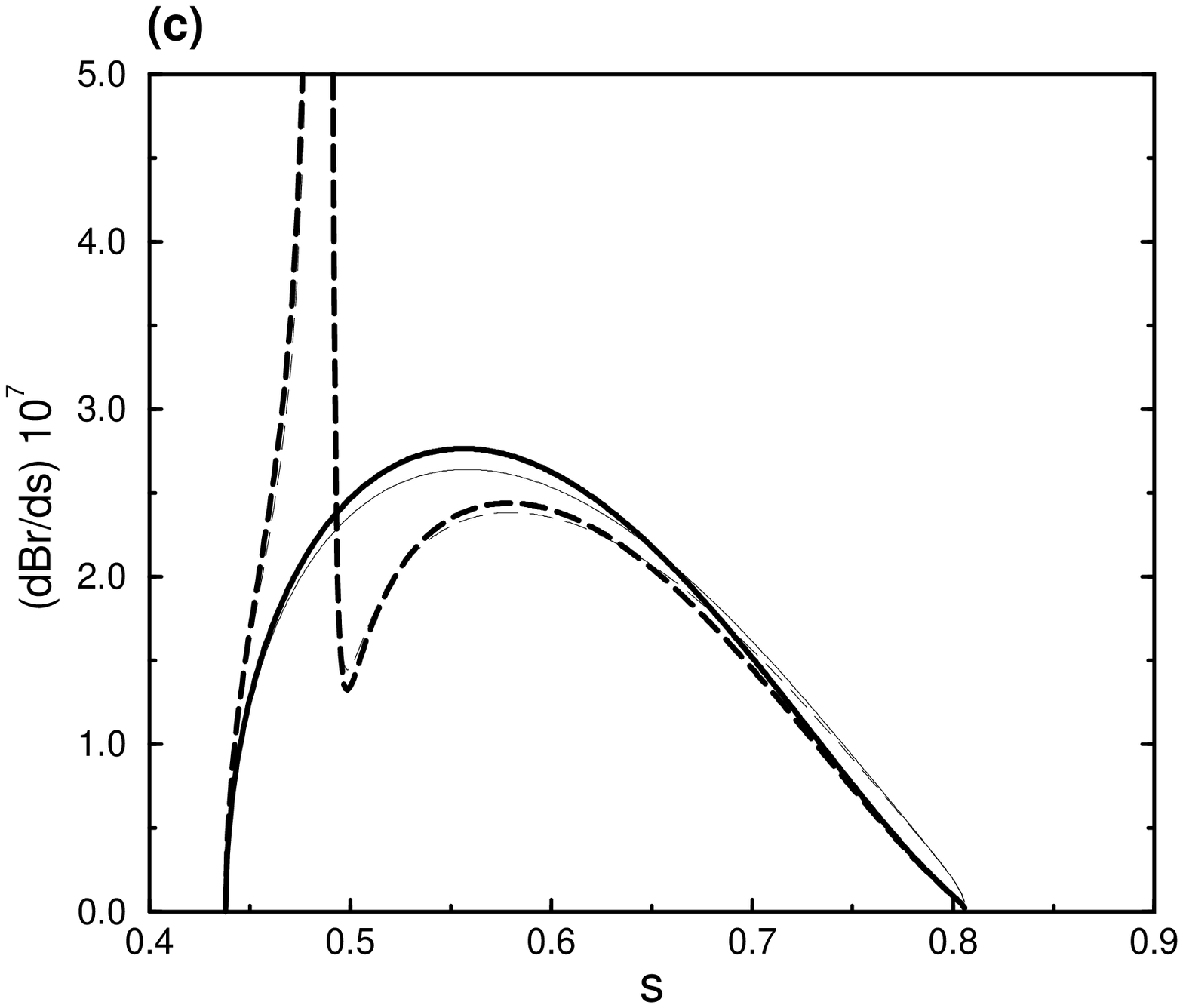}
\includegraphics{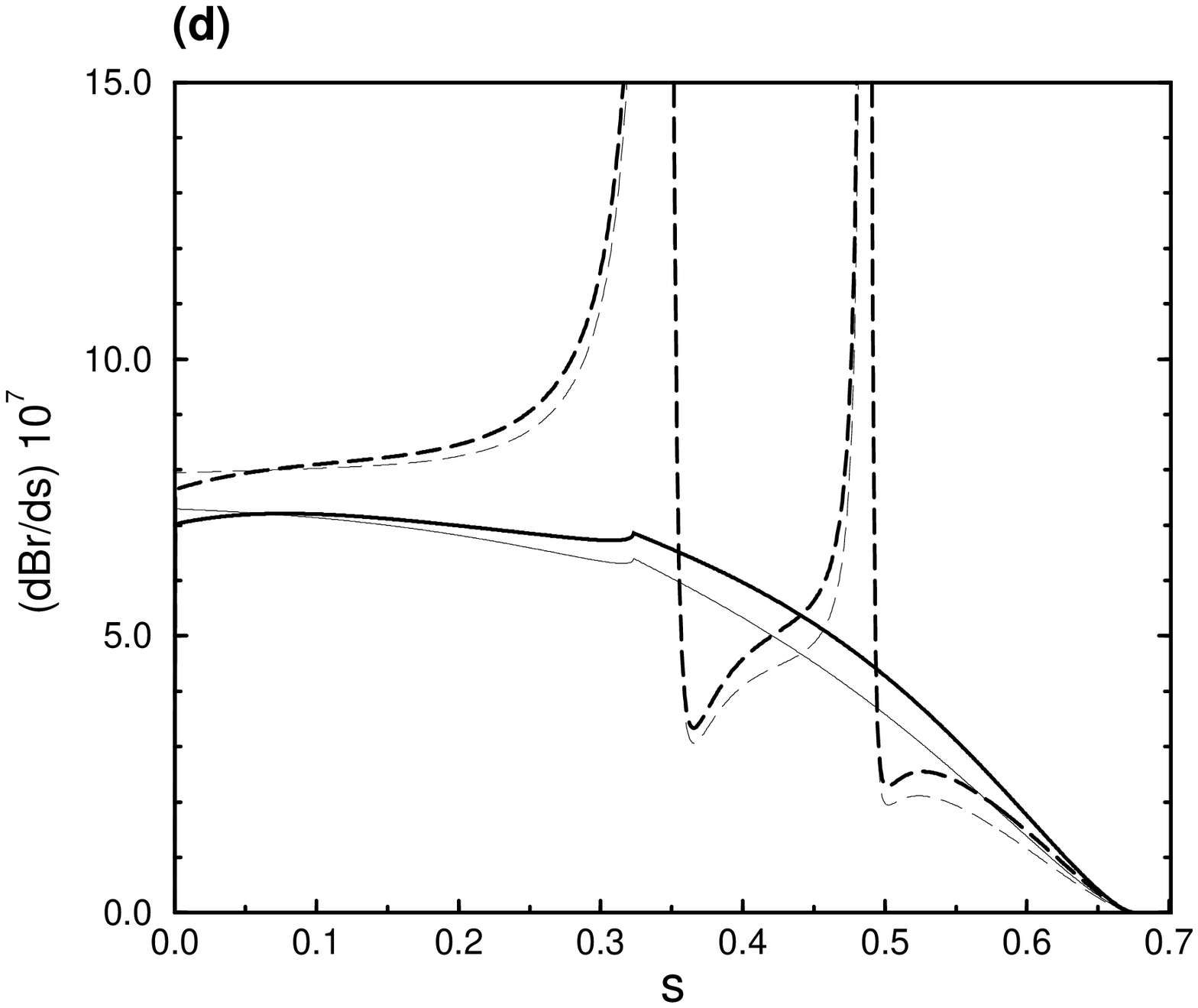}
\includegraphics{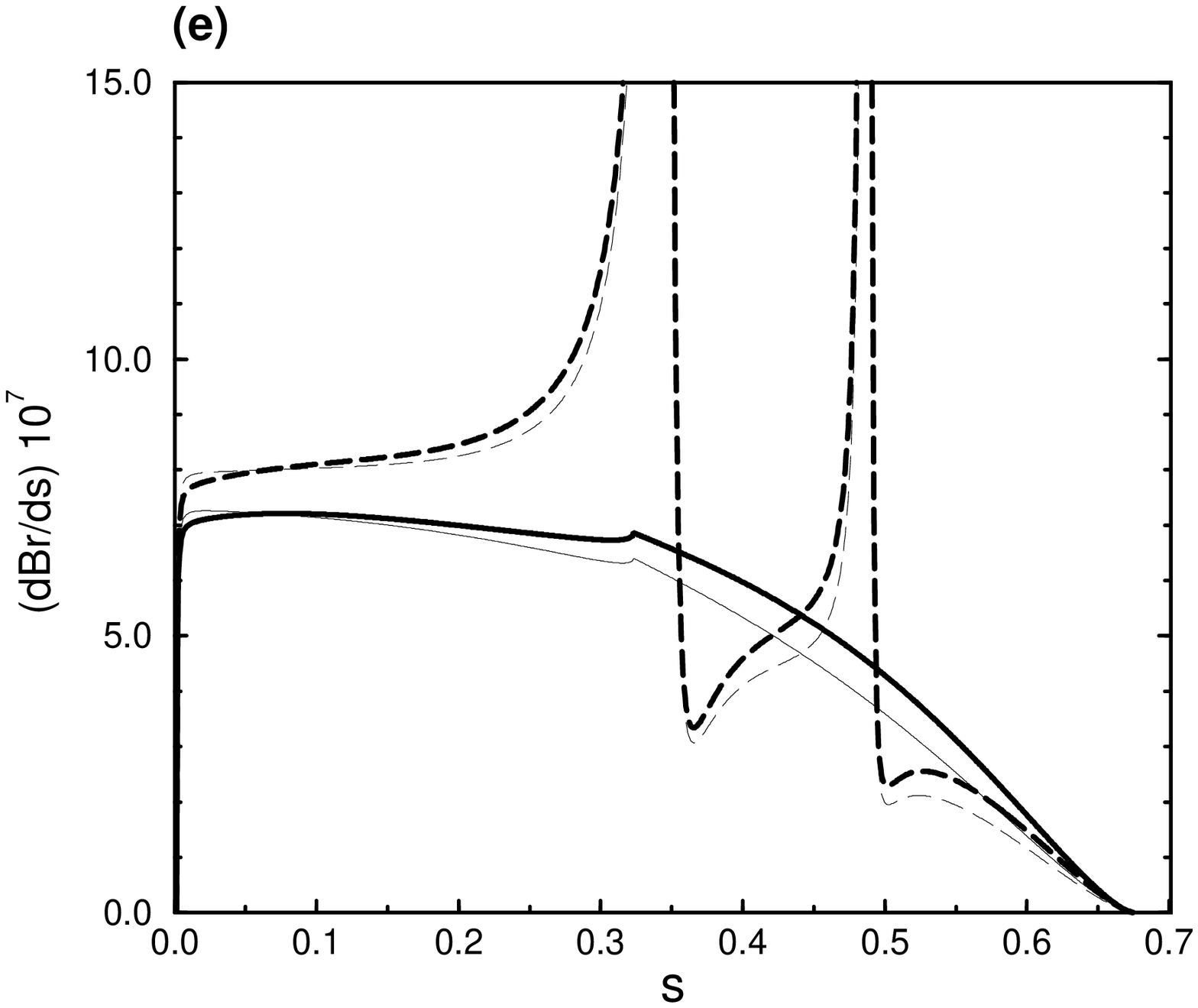}
\includegraphics{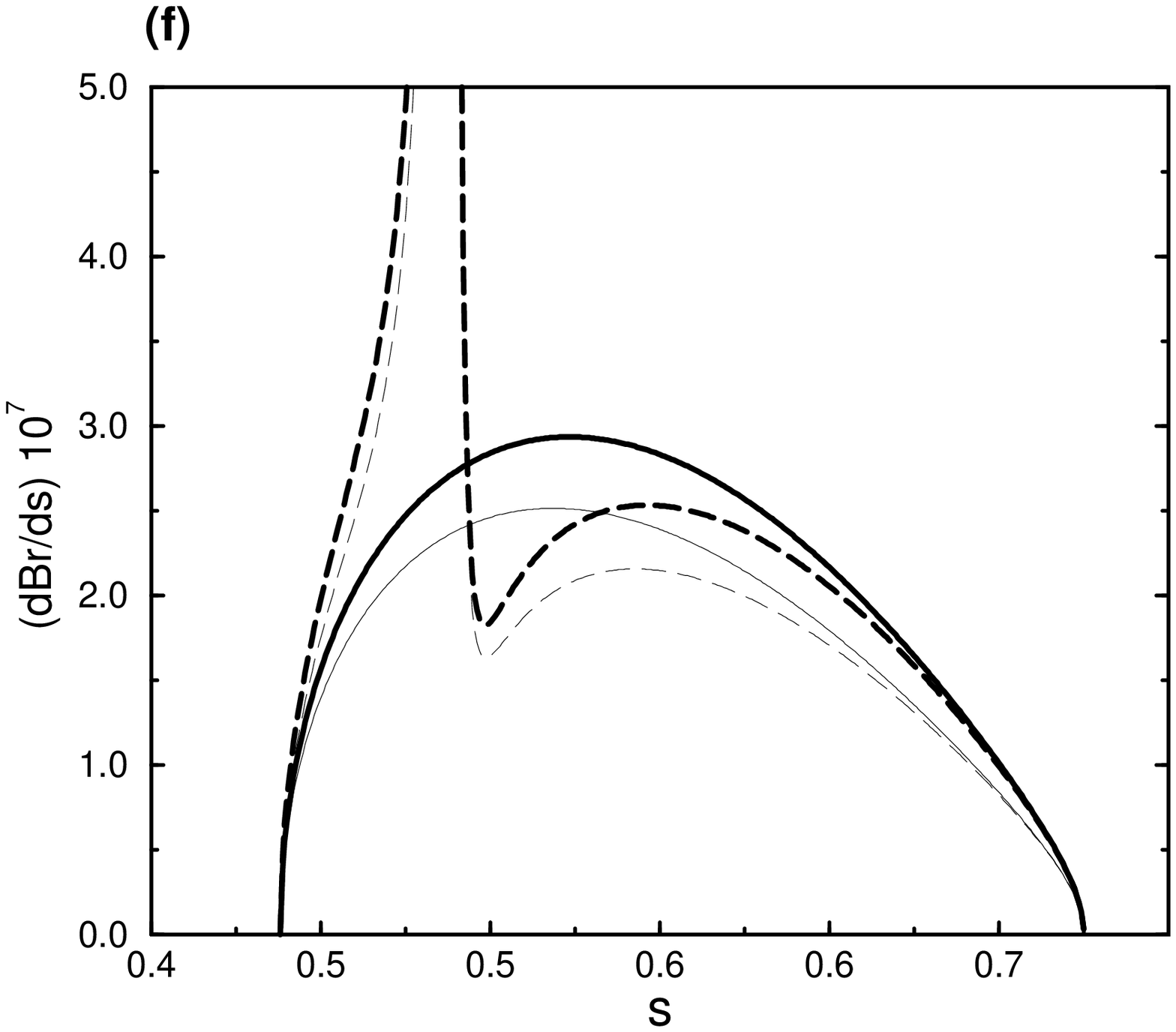}
\includegraphics{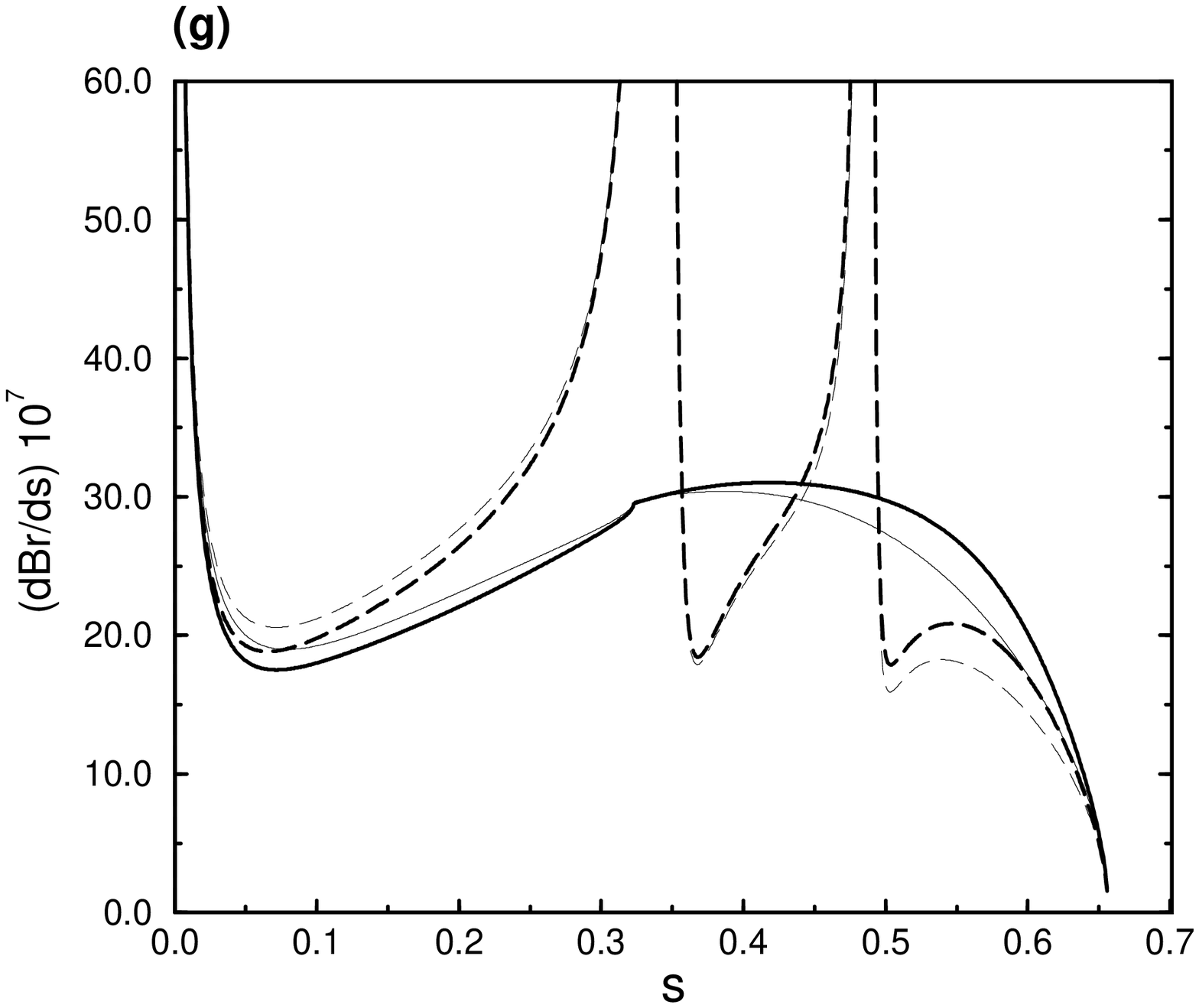}
\includegraphics{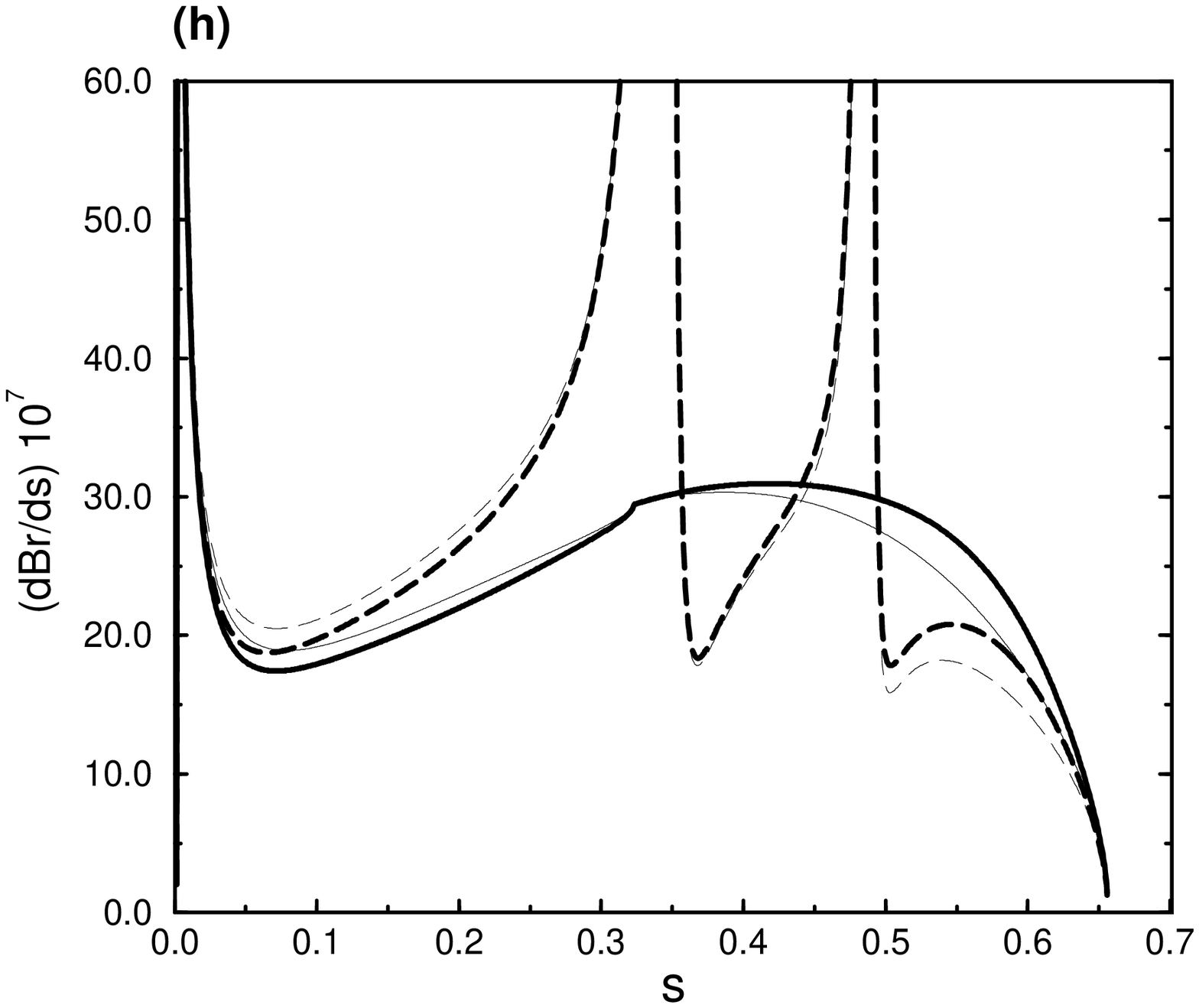}
\includegraphics{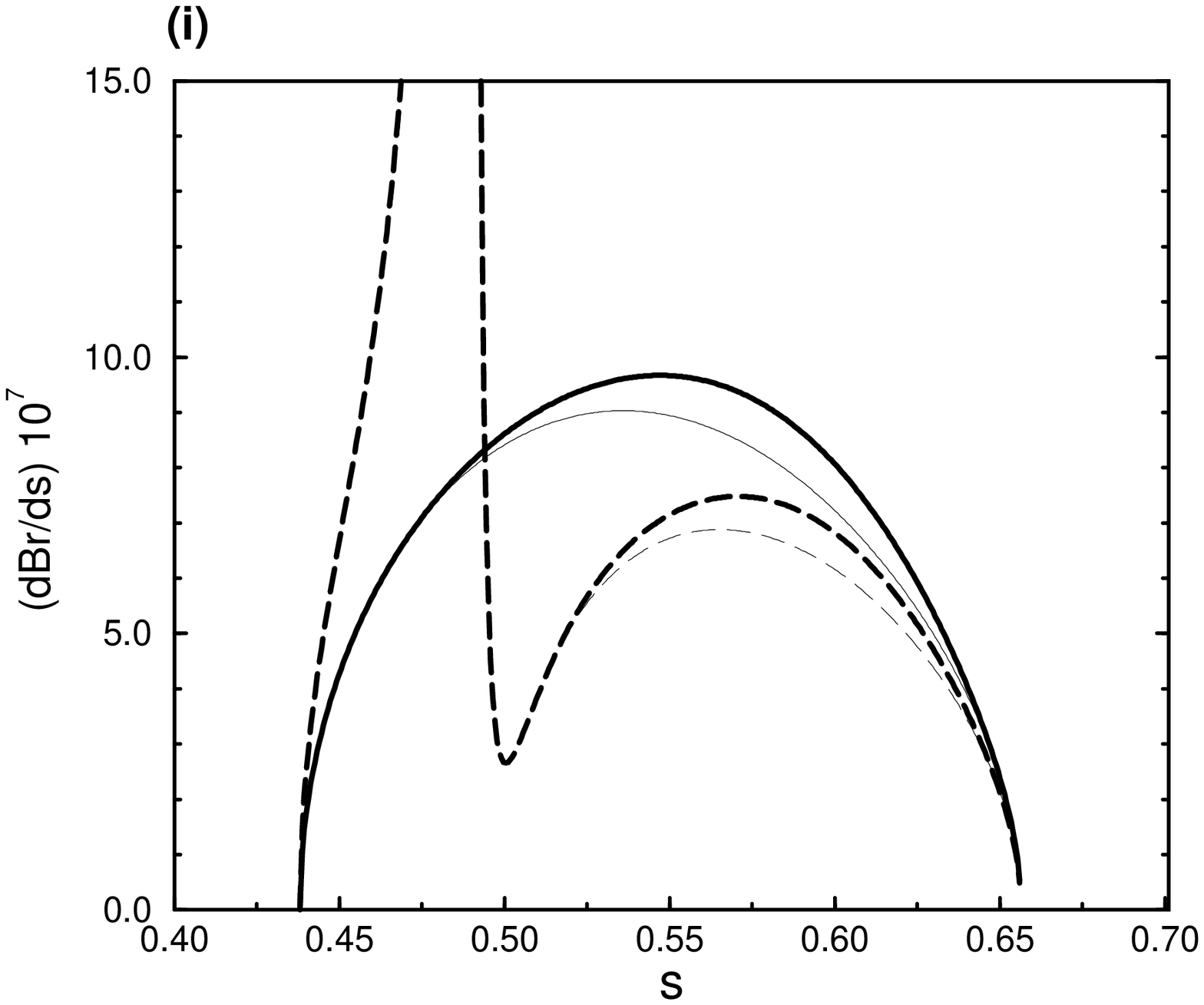}\vskip 17.0cm \caption{Differential decay
branching ratios for $B_s\to \eta \ell^+\ell^-$ in (a-c),
 $B_s\to\eta'\ell^+\ell^-$ in (d-f), and $B_s \to \phi
\ell^+\ell^-$ in (g-i) with $\ell$=e, $\mu$, and $\tau$,
respecitvely.
 The curves with (without)
resonant shapes represent including (non-including) LD
contributions.
 Legend is the same as Figure \ref{forms}.} \label{fig4}
\end{figure}

\begin{table}[h]
   \caption{Decay branching ratios of
$B_s \to (\eta, \eta', \phi) \ell^+\ell^-$ without including LD
effects.}
   \label{br2}
   \begin{center}
   \begin{tabular}{lccccc}
   \hline
   & \multicolumn{5}{c} {without LD}\\ \hline
   \multicolumn{1}{c}{Decay Mode} & LFQM & CQM & Ref. \cite{eta1} & Ref. \cite{phi1}&
   Ref. \cite{phi2} \\ \hline
   $10^{7}$Br$( B_s \to \eta e^+e^-) $
                      & $ 3.43 $ & $ 3.13 $ &-&-&-\\
   $10^{7}$Br$( B_s \to \eta \mu ^{+}\mu ^{-})$
                      & $ 3.42 $ & $ 3.12 $ &4.6,5.2&-&-\\
   $10^{7}$Br$( B_s \to \eta \tau ^{+}\tau ^{-}) $
                      & $ 0.67 $ & $ 0.67 $ &-&-&-\\
   $10^{7}$Br$( B_s \to \eta' e ^{+}e ^{-}) $
                      & $ 3.64 $ & $ 3.42 $&-&-&-\\
   $10^{7}$Br$( B_s \to \eta' \mu ^{+}\mu ^{-}) $
                      & $ 3.63 $ & $ 3.41 $&-&-&-\\
   $10^{7}$Br$( B_s \to \eta' \tau ^{+}\tau ^{-}) $
                      & $ 0.50 $ & $ 0.43 $&-&-&-\\
   $10^{6}$Br$( B_s \to \phi e ^{+}e ^{-}) $
                      & $ 1.72 $ & $ 1.69 $&-&-&2.01,1.87\\
   $10^{6}$Br$( B_s \to \phi \mu ^{+}\mu ^{-}) $
                      & $ 1.64 $ & $ 1.61 $&-&2.5&1.65,1.25\\
   $10^{7}$Br$( B_s \to \phi \tau ^{+}\tau ^{-}) $
                      & $ 1.60 $ & $ 1.51 $&-&-&1.38,2.28\\ \hline
   \end{tabular}
   \end{center}
\label{shortlepton}
\end{table}
By using the form factors of the LFQM and CQM in Figure
\ref{forms} and $|\lambda_t|=|V_{tb}V_{ts}|\simeq 0.041$, from
Eqs. (\ref{Rate1}) and (\ref{Rate2}), we now can estimate the
numerical values for the decay rates. In Figures \ref{fig3} and
\ref{fig4}, we show the differential decay branching ratios as a
function of $s=q^{2}/m_{B_s}^{2}$ for $B_s\to M_s\ell\bar{\ell}$
($\ell=\nu,e,\mu,\tau$) in the two models, respectively. In Tables
\ref{tablenu} and \ref{shortlepton}, we summarize the BRs of $B_s
\to (\eta, \eta', \phi)\ell^+\ell^-\ (l=\nu,e,\mu,\tau)$, where LD
effects for the charged lepton modes are not included. In Table
\ref{shortlepton}, we also show the results in Refs. \cite{eta1},
\cite{phi1} and \cite{phi2}.
In the table, the two values under Ref. \cite{eta1} represent two
inputs of the $\eta$ mixing parameter and those under Ref.
\cite{phi2}  are from two kinds of hadronic models.

With the LD effects, we need to introduce some cuts around the
resonances of $J/\psi$ and $\psi^{\prime }$. Explicitly, we use
the same cut as the one in the CDF experiment \cite{CDF:Bs}, which
leads to three regions given by
\be
   &I.& 2m_{\ell}< \sqrt{q^2}< 2.9; \nn \\
   &II.& 3.3\;<\;\sqrt{q^2}<3.6;\nn \\
   &III.& 3.8\;<\sqrt{q^2}<m_{B_s}-m_{M_s}.
\label{Cuts} \ee
In Table \ref{longlepton}, we present the decay
BRs in terms of the regions shown in Eq. (\ref{Cuts}).
  From Table \ref{longlepton}, it is interesting to pointed out that the results of
  $Br(B_s \to \phi\mu^+\mu^-)=1.23(1.24)\times 10^{-6}$ in the LFQM (CQM)
 are about two orders of magnitude less than the recent CDF experimental upper limit of
 $6.7\times10^{-5}$ \cite{CDF:Bs}.
 We note that with the LD effects, in Refs. \cite{phi1,phi2}  the branching ratio
of $ B_s \to \phi\mu^+\mu^-$ is found to be around $1.9$ and
$(1.91,1.45)\times 10^{-6}$, respectively. The values for Ref.
\cite{phi2} were derived with  a much smaller cut.

\begin{table}[h]
   \caption{Decay branching ratios of
$B_s \rightarrow
   (\eta, \eta', \phi)\ell^+\ell^-$ with LD effects and the cuts.}
   \label{br3}
   \begin{center}
   \begin{tabular}{|c|rr|rr|rr|rr|}
   \hline
   & \multicolumn{8}{c|} {with LD}\\ \hline
   regions & \multicolumn{2}{c|} {I}
           & \multicolumn{2}{c|} {II}
           & \multicolumn{2}{c|} {III}
           & \multicolumn{2}{c|} {I+II+III} \\ \hline
   Decay BR & LFQM & CQM & LFQM & CQM & LFQM & CQM & LFQM & CQM \\  \hline
     $10^{7}$Br$( B_s \to \eta e ^{+}e ^{-}) $
   & $1.86$ & $1.81$ & $0.32$ & $0.28$ & $0.61$ & $0.49$ & $2.79$& $2.59$ \\
     $10^{7}$Br$( B_s \to \eta \mu ^{+}\mu ^{-}) $
   & $1.85$ & $1.80$ & $0.32$ & $0.28$ & $0.61$ & $0.49$ & $2.78$& $2.58$ \\
     $10^{8}$Br$( B_s \to \eta \tau ^{+}\tau ^{-}) $
   & $--$ & $--$ & $0.13$ & $0.12$ & $4.95$ & $5.11$ & $5.08$& $5.23$ \\
     $10^{7}$Br$( B_s \to \eta' e ^{+}e ^{-}) $
   & $2.46$ & $2.42$ & $0.34$ & $0.30$ & $0.27$ & $0.22$ & $3.07$& $2.94$ \\
     $10^{7}$Br$( B_s \to \eta' \mu ^{+}\mu ^{-}) $
   & $2.44$ & $2.40$ & $0.35$ & $0.31$ & $0.27$ & $0.22$ & $3.05$& $2.92$ \\
     $10^{8}$Br$( B_s \to \eta' \tau ^{+}\tau ^{-}) $
   & $--$ & $--$ & $0.14$ & $0.13$ & $3.27$ & $2.77$ & $3.41$& $2.90$ \\
     $10^{7}$Br$( B_s \to \phi e ^{+}e ^{-}) $
   & $8.71$ & $9.04$ & $1.88$ & $1.83$ & $2.56$ & $2.23$ & $13.15$& $13.10$ \\
     $10^{7}$Br$( B_s \to \phi \mu ^{+}\mu ^{-}) $
   & $7.91$ & $8.30$ & $1.88$ & $1.83$ & $2.56$ & $2.23$ & $12.35$& $12.36$ \\
     $10^{8}$Br$( B_s \to \phi \tau ^{+}\tau ^{-})$
   & $--$ & $--$ & $0.48$ & $0.48$ & $8.87$ & $8.31$ & $9.35$& $8.80$ \\
   \hline
   \end{tabular}
   \end{center}
\label{longlepton}
\end{table}

\subsection{Asymmetries}
\begin{figure}[h]
\includegraphics{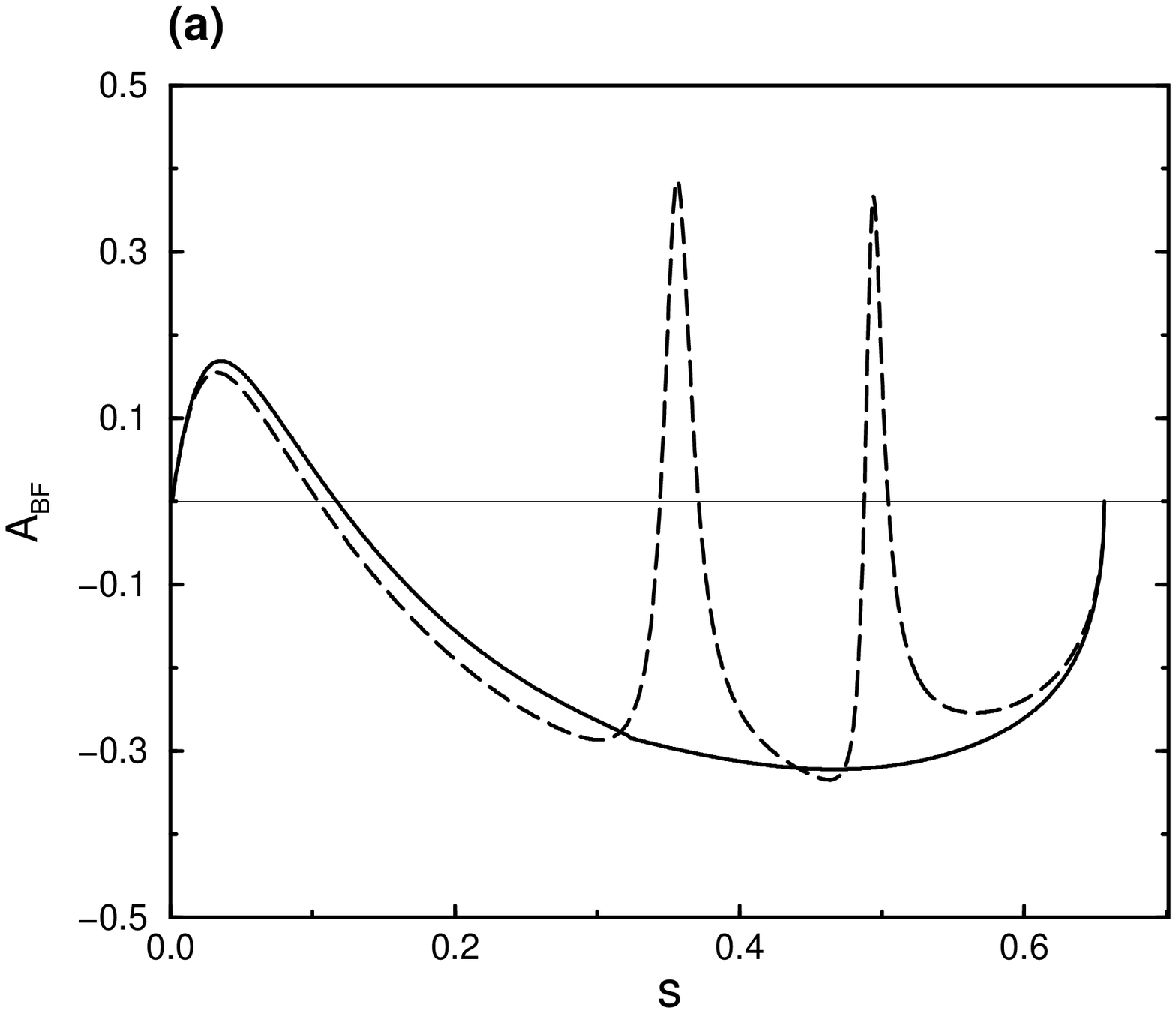}
\includegraphics{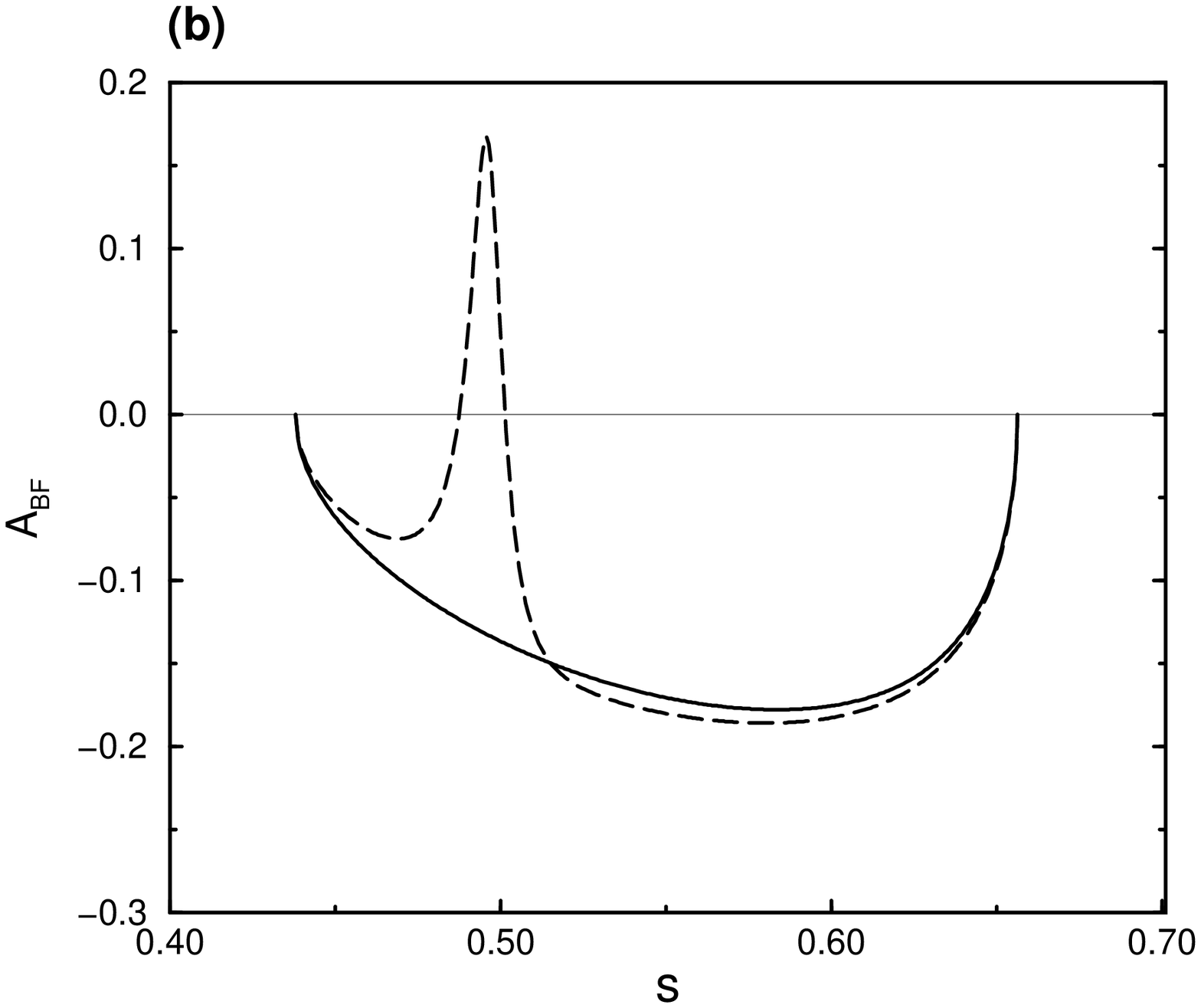}
\vskip 12.0cm \caption{ Forward-backward asymmetries in the LFQM
for (a) $B_s\to \phi \mu^+\mu^-$ and (b) $B_s\to \phi
\tau^+\tau^-$. Legend is the same as Figure \ref{fig4}.
}
\label{asym:afb}
\end{figure}

The forward-backward asymmetry (FBA) in the charged lepton decay
is defined by
\be
  {\cal A}_{FB}=\frac{\int_0^1d(\cos \theta_\ell)d\Gamma-\int_{-1}^0d(\cos
  \theta_\ell)d\Gamma}{\int_0^1d(\cos \theta_\ell)d\Gamma+\int_{-1}^0d(\cos
  \theta_\ell)d\Gamma}
\ee where $\theta_\ell$ is the angle between the lepton $\ell^+$
momentum in the dilepton c.m. frame and the $B_s$ meson decay l.b.
frame. In the SM, it is easy to show that the FBAs in $B_s\to
(\eta,\eta')\ell^+ \ell^-$ are equal to zero and the FBA in
$B_s\to \phi \ell^+ \ell^-$ is given by
\be
     {\cal A}_{FB}=\frac{12s
  \varphi_{\phi}^{1\over2}\left(1-\frac{4t}{s}\right)^{1\over2}(1-r_{\phi})~Re(G_V
     F_A^{0*}+G_A^0F_V^*)}
     {(1+\frac{2t}{s})~\alpha_{\phi}+t~\delta_{\phi}}\,.
\ee
 In Figure \ref{asym:afb}, we display ${\cal A}_{FB}(s)$ in
$B_s\to \phi \ell^+ \ell^-$ $(\ell=\mu,\tau)$ as a function of
$s$. As seen from the figure,  ${\cal A}_{FB}(B_s\to \phi
\mu^+\mu^-)$ changes sign at $s=s_0\simeq 0.1$, i.e., ${\cal
A}_{FB}(s_0)=0$, where $s_0$ satisfies the relation
\be
    s_0=\frac{C_7}{C_8^{eff}}\hat{m}_b(1+\sqrt{r_{\phi}})({g\over V}+{a_0\over
    A_0}).
\label{s0} \ee

Similar to  $B\to K^*\ell^+ \ell^-$ with $\ell=e$ or $\mu$, the
location of $s_0$ depends on the WCs and is insensitive to the
hadronic uncertainties \cite{Ali,burdman1}.

By writing the unit vector of
$\hat{n}={\vec{p}_\ell \over |\vec{p}_\ell|}=\pm1$,
we can define the longitudinal lepton polarization asymmetry in
 $B_s\to M_s\ell^+\ell^-$ as follows
\be
   P_L(s)=\frac{~~{d\Gamma(\hat{n}=-1)\over ds}~-~
                  {d\Gamma(\hat{n}=1)\over ds}~~}
               {~~{d\Gamma(\hat{n}=-1)\over ds}~+~
                  {d\Gamma(\hat{n}=1)\over ds}~~}\,.
\label{PL}
 \ee
From Eqs. (\ref{Metall}), (\ref{Mphill}) and (\ref{PL}),
$P_L$ in
 $B_s\to P\ell^+\ell^-\ (P=\eta,\eta')$ and $B_s\to \phi\ell^+\ell^-$
are given by
\be
     P_L^{P} &=& (C^{P}_\theta)^2\times \frac{2\left( 1-\frac{4t}{s}\right)
     ^{1\over2}} {\left(1+\frac{2t}{s}\right)\alpha_{P}
     +t ~\delta_{P}}
     ~Re~\left[\varphi_{P}\left(C_8^{eff}F_+
     -2\frac{C_7F_T}{1+\sqrt{r_{P}}}
     \right)(C_9F_+)^*\right]\,\,
\\
     P_L^{\phi} &=&\frac{2\left( 1-\frac{4t}{s}\right)^{1\over2}}
     {\left(1+\frac{2t}{s}\right)\alpha_{\phi} +t ~\delta_{\phi}}
     ~Re~\Big\{\frac{1}{r_{\phi}}\left[\varphi_{\phi}^2~G_A^+F_A^{+*}
     +\varphi_{\phi} (1-r_{\phi})(1-r_{\phi}-s)
      \right. \nn \\
     && \left.
     \left(G_A^0F_A^{+*}+G_A^+F_A^{0*}\right)
     +(\varphi_{\phi}+12r_{\phi}~s)(1-r_{\phi})^2~G_A^0F_A^{0*}
     +8\varphi_{\phi} r_{\phi}~s~G_VF_V^*\right]
     \Big\}\,,
\label{PLphi}
 \ee respectively, where $C^\eta_\theta=-\sin\theta$
and $C^{\eta'}_\theta=\cos\theta$.

\begin{figure}[h]
\includegraphics{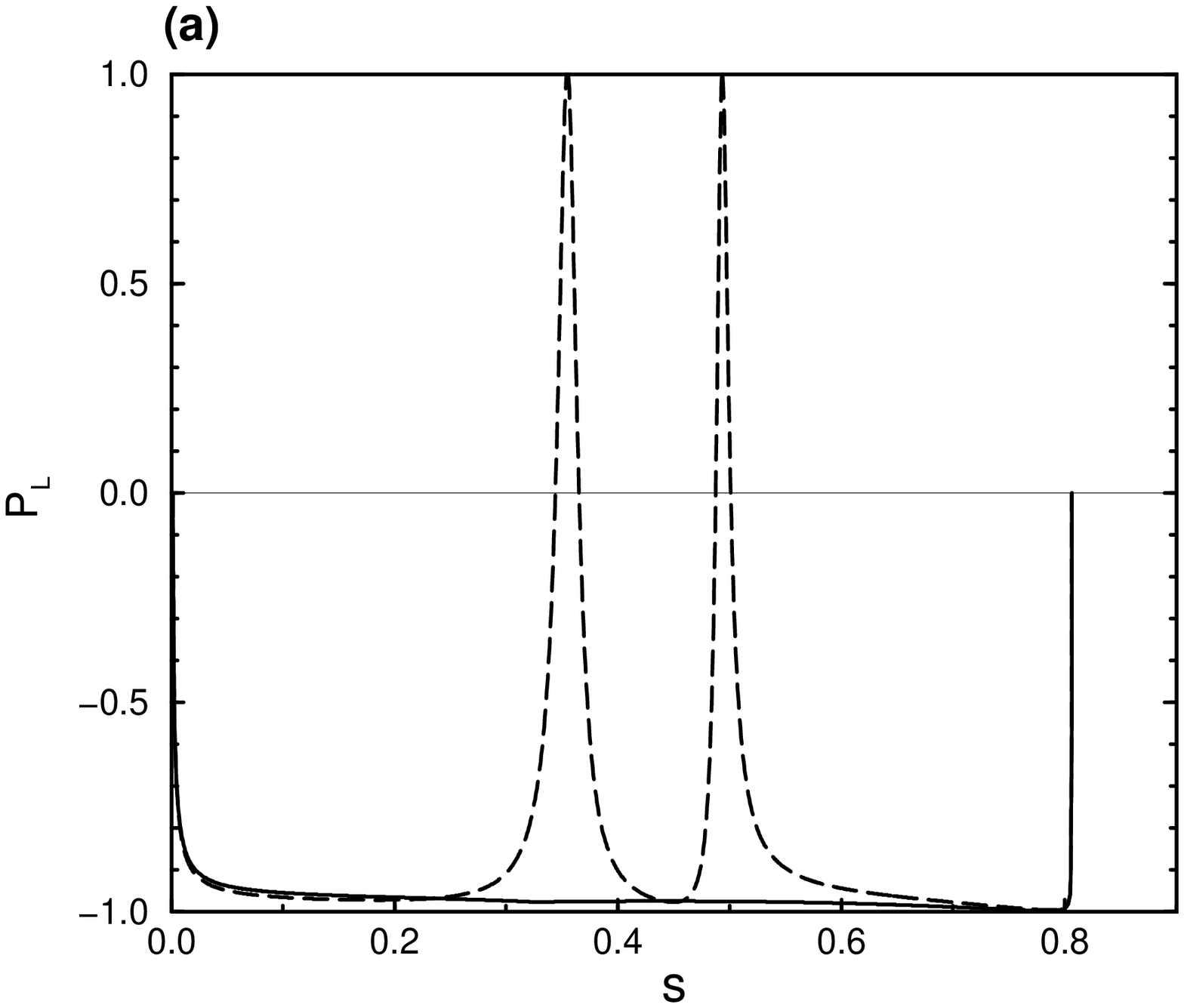}
\includegraphics{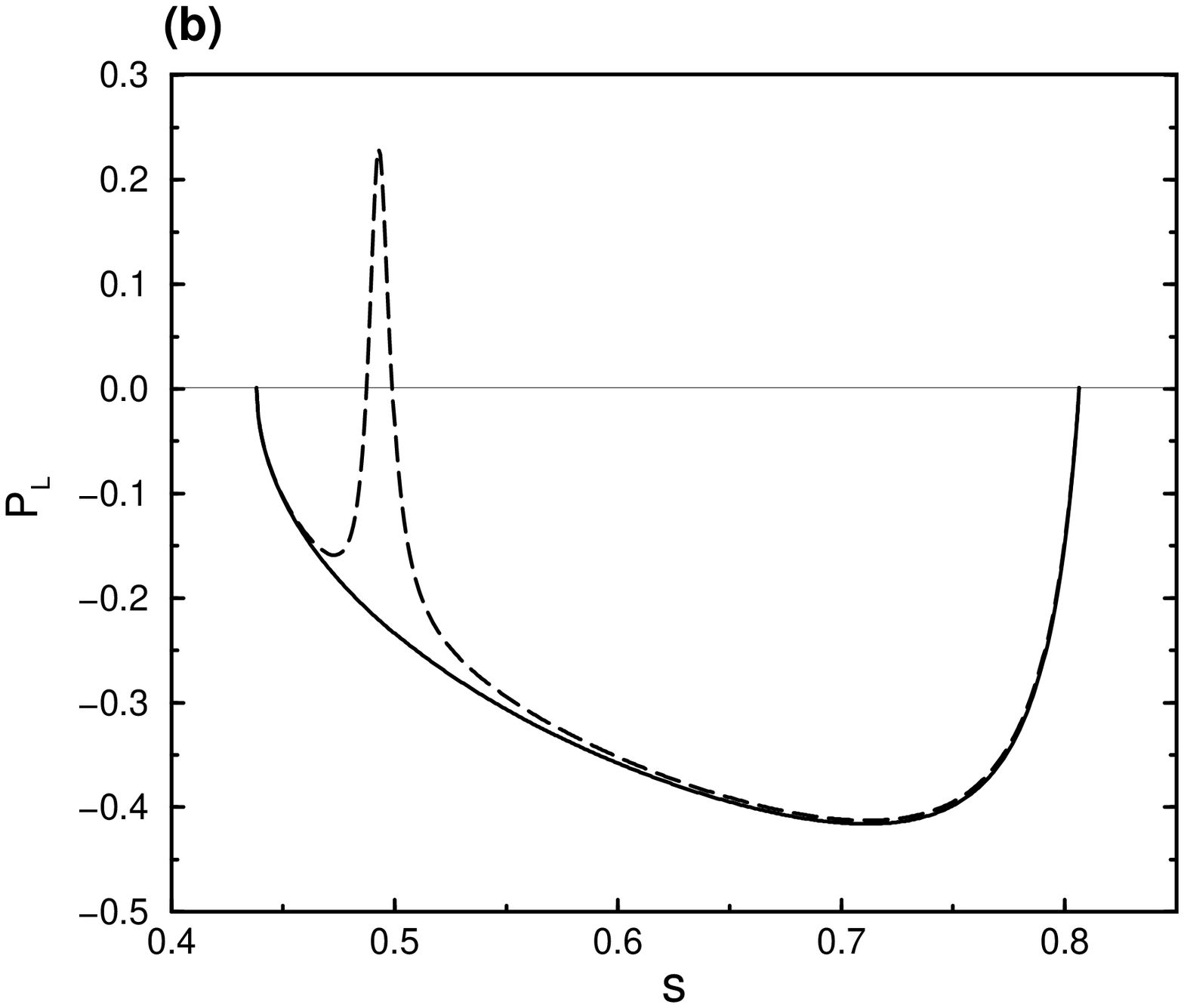}
\includegraphics{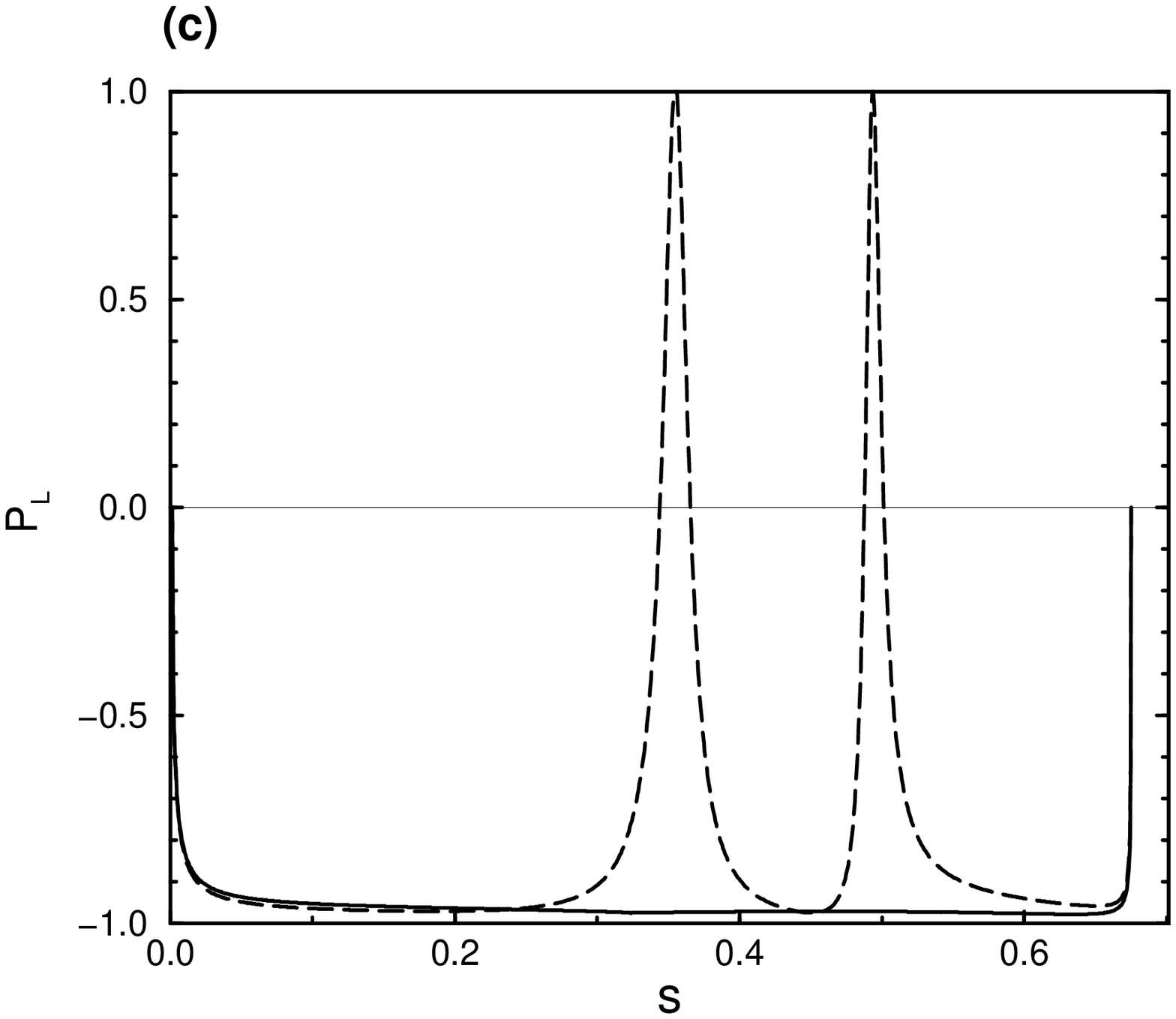}
\includegraphics{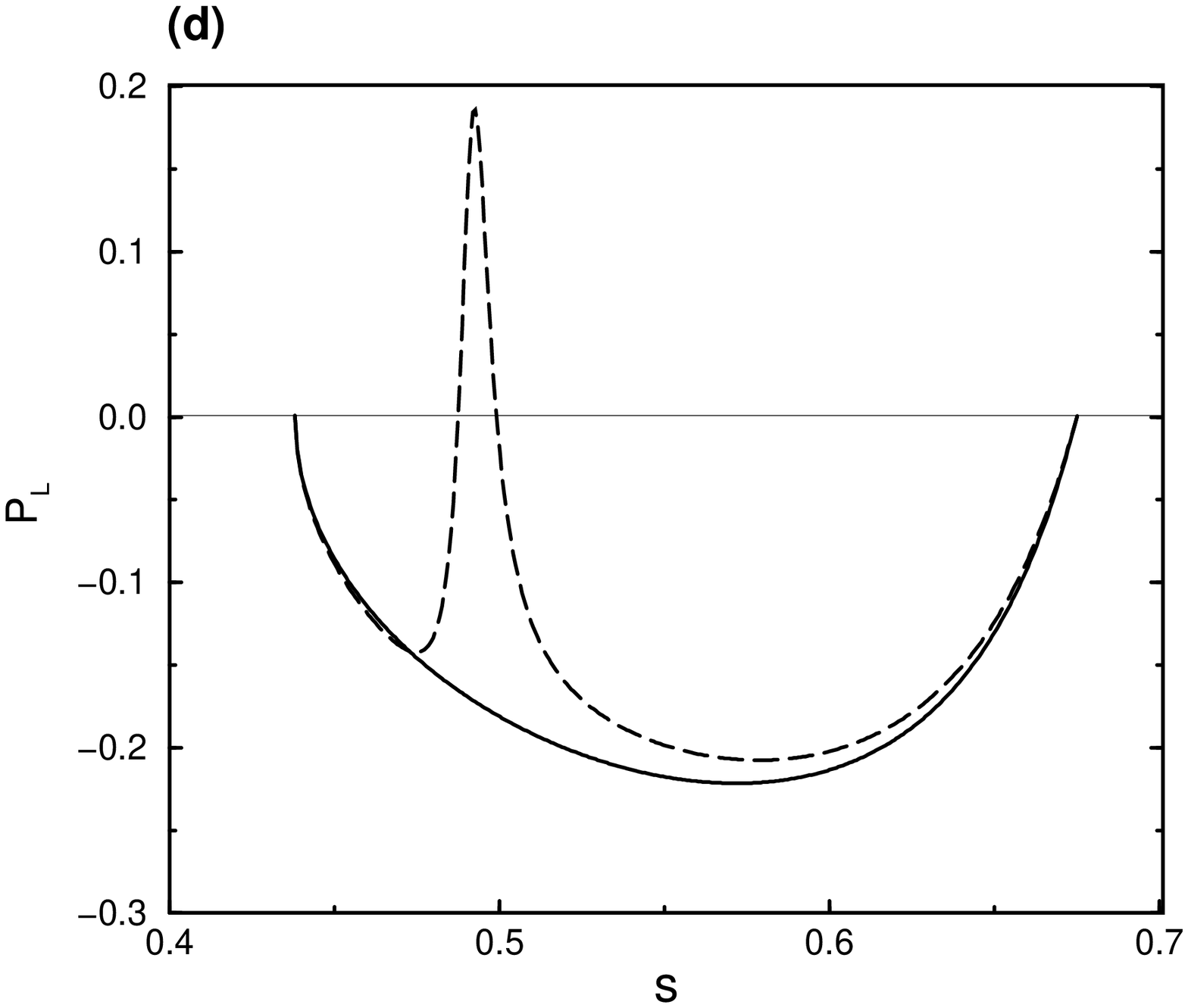}
\includegraphics{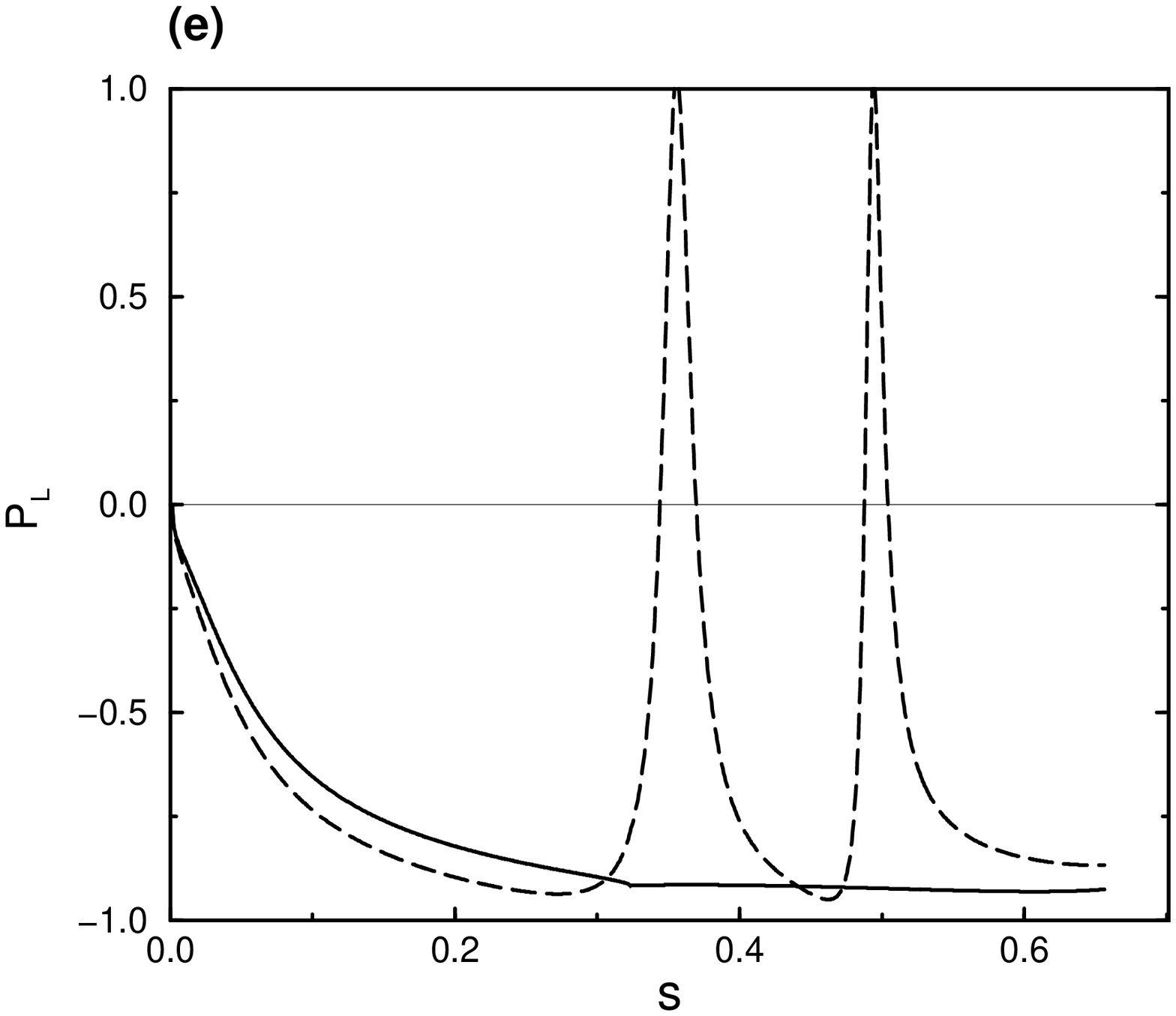}
\includegraphics{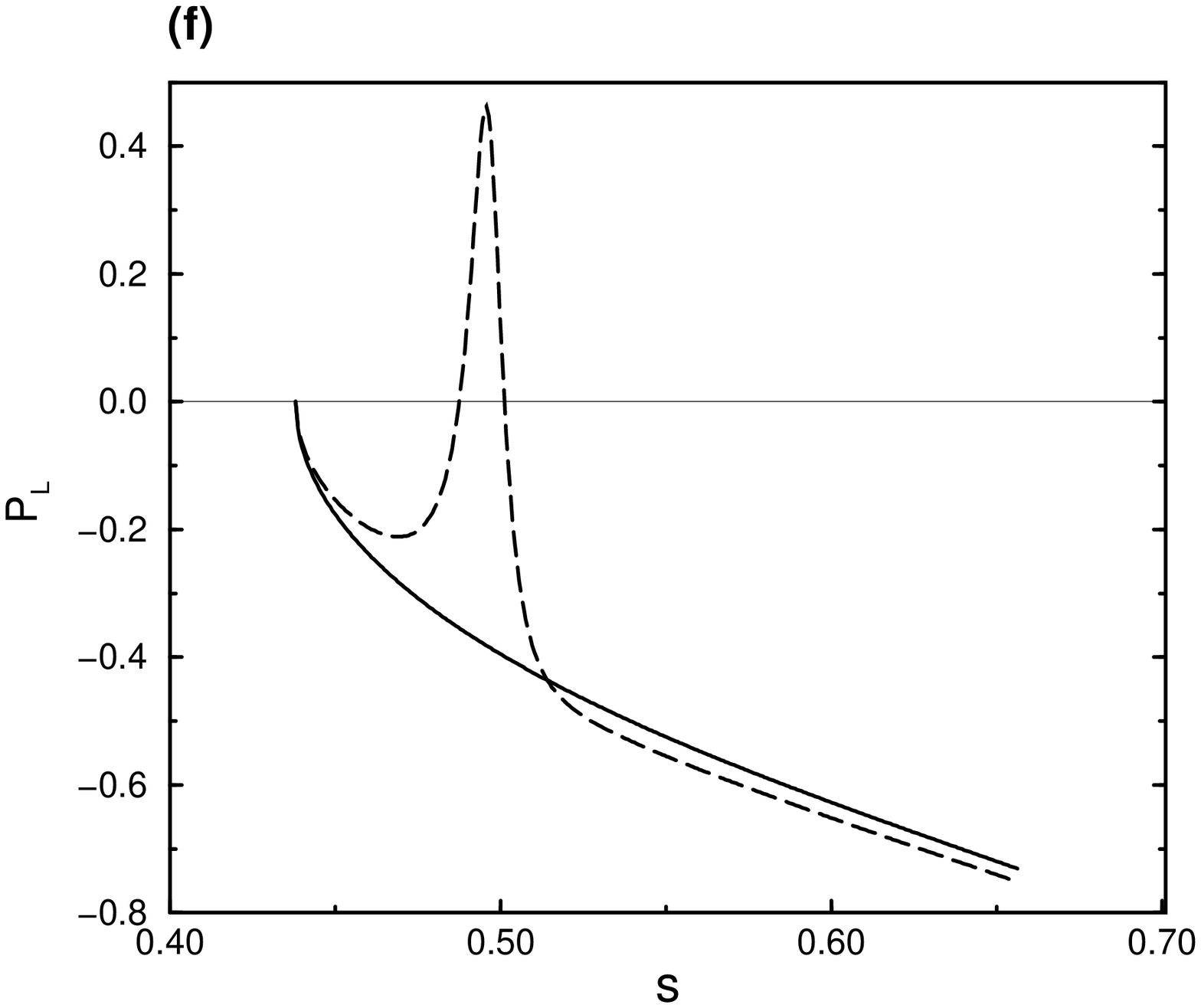}
\vskip 17.0cm \caption{ Longitudinal lepton polarization
asymmetries in the LFQM for $B_s\to \eta \ell^+\ell^-$ in (a,b),
$B_s\to\eta'\ell^+\ell^-$ in (c,d), and $B_s \to \phi
\ell^+\ell^-$ in (e,f) with $\ell$=($\mu,\tau$). Legend is the
same as Figure \ref{asym:afb}.
} \label{asym:pl}
\end{figure}

In Figure \ref{asym:pl}, we show the longitudinal lepton
polarizations in $B_s\to (\eta,\eta',\phi)\ell^+\ell^-$ with
$\ell=\mu$ and $\tau$. The results for the electron modes are
similar to the muon ones. We remark that, as shown in the figures,
$P_L^{\eta,\eta'}(s)$  are close to -1 for the muon modes except
the end points of $q^2_{min}=4m_{\ell}^2$ and
$q^2_{max}=(m_{B_s}-m_M)^2$ in which they are zero and they range
from $-0.4$ to $0$ and $-0.2$ to $0$ for the $\tau$ ones,  while
the average ones of $<P_L^{\eta(\eta')}>$ without LD effects are
$-0.97 (-0.96)$ and $-0.32 (-0.17)$ for the muon and tau modes,
respectively. For $B_s\to\phi\ell^+\ell^-$,
$P_L^{\phi}(s_{min})=0$ but $P_L^{\phi}(s_{max})\neq 0$ since
$\varphi_{\phi}$ cannot be factorized out in Eq. (\ref{PLphi}). We
have that $<P_L^{\phi}>_{\mu}=-0.81$ and
$<P_L^{\phi}>_{\tau}=-0.49$.


\section{Conclusions}
We have studied the rare $B_s$ decays of $B_s \to
(\eta,\eta',\phi) \nu \bar{\nu}$ and $B_s \to
(\eta,\eta',\phi)\ell^+\ell^-$ ($\ell=e,\mu,\tau$). In our
analysis, we have used the form factors of
 $B_s \to (\eta,\eta',\phi)$ transitions calculated in
the LFQM and CQM. We have found that $Br(B_s \to \eta
\ell\,\bar{\ell})\ (\ell=\nu,e,\mu,\tau)$= $(23.4, 3.43, 3.42,
0.67)$ and $(21.7, 3.13, 3.12, 0.67)\times 10^{-7}$, $Br(B_s \to
\eta' \ell\,\bar{\ell})$=$(25.2, 3.64, 3.63, 0.50)$
 and $(23.8, 3.42, 3.41, 0.43)\times 10^{-7}$, and
$Br(B_s \to \phi \ell\,\bar{\ell})$= $(12.02, 1.72, 1.64, 0.16)$
and $(11.65, 1.69, 1.61, 0.15)\times 10^{-6}$ without LD
contributions in the two models, respectively. We have also
discussed the longitudinal lepton polarization and
forward-backward asymmetries in the charged lepton modes. We have
shown that the behaves of the asymmetries are
 similar to those in $B\to K^{(*)}\ell^+\ell^-$.
 Clearly, some of the above rare $B_s$ decays and asymmetries can
 be measured at the BTeV and LHC-B experiments.
\\

\section*{Acknowledgments}
This work was supported in part by the National Science Council of
the Republic of China under contract numbers
 NSC-91-2112-M-007-043.


\newpage

\newpage

\newpage

\newpage

\newpage

\newpage


\end{document}